\documentclass{aa}  

\usepackage{graphicx}
\usepackage{natbib}
\usepackage{color}
\usepackage{txfonts}
\usepackage{bm} 
\usepackage[colorlinks=true,linkcolor=blue,citecolor=blue,urlcolor=blue]{hyperref}
\begin{document}

    \title{Dusty disks as safe havens for terrestrial planets: Effect of the back-reaction of solid material on gas}
    \titlerunning{Dusty disks as safe havens for terrestrial planets}
    \authorrunning{Regály et al.}

   \author{Zs. Regály, 
          \inst{1}\fnmsep\inst{2}\thanks{E-mail: \href{regaly@konkoly.hu}{regaly@konkoly.hu}}
          A. Németh,
          \inst{1}\fnmsep\inst{2}\fnmsep\inst{3}
          G. Krupánszky,
          \inst{1}\fnmsep\inst{2}\fnmsep\inst{3}
          and Zs. Sándor
          \inst{3}\fnmsep\inst{1}\fnmsep\inst{2}
          }

   \institute{
   Konkoly Observatory, HUN-REN, Research Centre for Astronomy and Earth Sciences, \\ 
   H-1121 Budapest, Konkoly Thege Miklós út 15-17, Hungary.
         \and
         CSFK, MTA Centre of Excellence, Konkoly Thege 15-17, 1121, Budapest, Hungary 
         \and
             ELTE Eötvös Loránd University, Institute of Physics and Astronomy, Department of Astronomy, \\
             H-1117 Budapest, Pázmány Péter sétány 1/A, Hungary.
             }

\date{}
 
  \abstract
   {Previous studies have shown that there is considerable variation in the dust-to-gas density ratio in the vicinity of low-mass planets undergoing growth. 
   {This can lead to a significant change in the planetary momentum exerted by the gas and solid material.
   However, due to the low dust-to-gas mass ratio of protoplanetary disks (about one percent), the effect of the solid material on the gas dynamics -- that is, the back-reaction of the solid material -- is often neglected. }
   }
   {We aim to study the effect of the back-reaction of solid material on the torques felt by {low-mass} planets.
   The effect of the back-reaction of solid material is investigated by comparing non-accreting and accreting models.}
   {We performed locally isothermal, global two-dimensional hydrodynamic simulations of planet-disk {interactions} using the code {\small GFARGO2}. 
   Low-mass planets in the range of $0.1-10~M_\oplus$ {accrete only} solid material.
   The solid component of the disk was {treated} as a pressureless fluid.
   Simulations were compared with taking and not taking into account the back-reaction of the solid material on the gas.
   {The solid component was assumed to have a fixed Stokes number in the range of $0.01-10$.
   All models assumed a canonical solid-to-gas mass ratio of 0.01.}
   }
  {{The back-reaction of the solid has been shown to have a significant effect on the total torque exerted on a low-mass planet.}
  In general, the inclusion of the back-reaction results in a {greater number of models with positive torque values} compared to models that neglect the back-reaction.
  {It is clear, therefore, that the simulation of planetary growth and migration via hydrodynamic modeling requires the inclusion of a solid-gas back-reaction.}
  As a {result} of the back-reaction and accretion, a Mars-sized planetary embryo will experience positive total torques from the disk containing coupled solid components ($\mathrm{St}\leq0.01$).
  Earth-mass planets also experience positive total torques from the disk containing boulder-sized solid components ($2\leq\mathrm{St}\leq5$).
  The accretion of weakly coupled solid material tends to increase the positive torques and decrease the negative torques.
   }
   {
   Our {results} suggest that the combined effect of back-reaction and accretion is beneficial to the formation of planetary systems by reducing the {likelihood} of a young planet being engulfed by the central star.
   }

   \keywords{accretion, accretion disks 
        --- hydrodynamics 
        --- methods: numerical 
        --- protoplanetary disks 
        --- planet-disk interactions
               }

   \maketitle
%

\section{Introduction}




Planets and massive planetary cores embedded in protoplanetary disks change their orbital elements due to the torques exerted by the {surrounding} disk. 
These torques {typically} alter the semi-major axes of planetary objects {and also effectively} damp their eccentricities and inclinations. 
If the net torque exerted by the planet is negative, {as in the case of} locally isothermal disks \citep{Tanaka+2002ApJ}, the planet could be lost to the planet-forming processes by a {rapid} inward migration. 
Several mechanisms against inward migration have been considered: density maxima \citep{Lyra+2008,Sandor+2011ApJ,GuileraSandor2017A&A}, sudden changes in the thermal property from adiabatic and isothermal states \citep{Lyra+2010ApJ}, the heating force due to the luminosity of the pebble-accreting planet \citep{Benitez2015}, and the back-reaction of the solid material to the planet accumulating around it \citep{BenitezPessah2018,Regaly2020, Guileraetal2023, Chrenko2024A&A...690A..41C}. 
All these mechanisms {act} against inward migration, allowing sufficient time for the planetary core to accrete solid material and eventually collect a gaseous envelope to become a giant planet. In this paper, the effect of the torque resulting from the back-reaction of solid material on the planet is {further investigated}, based on the work of \cite{Regaly2020}.

It is {known} that the solid-to-gas mass ratio in protoplanetary disks is about one percent (see, e.g., \citealp{WilliamsCieza2011}). 
Consequently, it is {reasonable to assume} that the gravitational influence of dust is insignificant {compared} to that of gas. 
Nevertheless, the asymmetry in the distribution of solid matter in the vicinity of the planet {can lead to a significant change} in the torque experienced by the planet. 
This phenomenon {is similar to} the heating torque induced by the asymmetric distribution of gas within the planetary Hill sphere.
Indeed, the combined effect of the solid-gas interaction and the gravitational pull of the planet {can lead to} the formation of a highly asymmetric distribution of solid material in the vicinity of low-mass planets. 

{It has been shown} by \citet{BenitezPessah2018} that the spatial distribution of solid material can be markedly asymmetric, such that it modifies the velocity or even reverses the direction of type I migration.
{However, planetary accretion} modifies the spatial distribution of solid material by {removing} solid material from the vicinity of the planet. 
In \citet{Regaly2020}, a detailed analysis is presented of how planetary accretion of solid material modifies the spatial distribution of solid species {of different sizes}, and thus the solid torque felt by low-mass planets.
As a consequence of solid accretion, the spatial asymmetry that has developed in the solid distribution in the vicinity of the planet becomes more pronounced.
It is {shown} that the magnitude of the solid torques increases with the strength of the accretion process. 
In the case of {low-mass} planets ($\leq1~M_\oplus$), the magnitude of the solid torque can {exceed} that of the gas, resulting in either a total torque reversal (even in locally isothermal disks) or {an enhanced} negative total torque.

The influence of the back-reaction of solid material on gas dynamics (hereafter, the term “feedback” is used) is {often neglected} in numerical simulations modeling planet-disk interactions (see, e.g., \citealp{BenitezPessah2018,Regaly2020,Chrenko2024A&A...690A..41C}).
However, several studies have already {alerted us to} the importance of dust feedback (see, e.g., a summary in \citealp{Gonzalezetal2018}).
Two-dimensional two-fluid hydrodynamic simulations by \citet{Kanagawaetal2017} show that if the grains are sufficiently large or piled up, the back-reaction is so effective that it forces the gaseous component of the disk to flow outward.
{In addition}, simulations of multiple populations of solid species in the disk {performed} by \citet{Dipierroetal2018} {find that} the back-reaction of solid material {reduces} the gas accretion {flux} to the central star compared to the solid-free case.
Moreover, in the outer disk, where the Stokes number can be large, the back-reaction can even drive the {gas flow outward}.
\citet{Gonzalezetal2017} show that with dust back-reaction, a self-induced dust trap develops in the disk, allowing grains to overcome the fragmentation barrier, and the largest grains to overcome the radial drift barrier.
High-resolution two-dimensional numerical simulations of non-accreting $2~M_\oplus$ {planets} by \citet{HsiehLin2020} show that inward migration can be slowed down by dusty dynamical {corotational} torques in inviscid disks.
The above studies show that the back-reaction of solid species can play {an important} role in planet formation and cannot be ignored in models of planet-disk interactions.

In this paper, we build on our previous analysis \citep{Regaly2020} of the torques exerted on low-mass planets that are accreting solid material by incorporating the effect of the back-reaction of solid material. 
Two-dimensional hydrodynamic simulations are {performed} with and without the effect of back-reaction, using non-accreting and accreting models, to elucidate the significance of back-reaction in the study of planet-disk interactions.

The following is a description of the structure of the paper: {Section~2 presents} the hydrodynamical model {used} and the initial conditions of the gas and solid material. Section~3 presents the impact of back-reaction on an Earth-mass planet, as well as the results of our parameter study, in which we vary the mass, the surface mass density profile of the disk, and the accretion rate.
Section~4 presents a discussion of the solid and gas torque profiles of different solid species in the {vicinity} of a low-mass planet, assuming different accretion rates.
The limitations of our model are also {discussed} in Sect.~4. 
The paper concludes with a summary of our results in Sect.~5. 
{Additional} torque profiles and material distributions are presented in Appendices~A and B.

\section{Hydrodynamical model}
\subsection{Governing equations}

We performed global two-dimensional hydrodynamic simulations of {planet-disk interactions} using the code {\small GFARGO2} \citep{RegalyVorobyov2017,Regaly2020}, a {GPU-enabled} version of the {\small FARGO} code \citep{Masset2000}.
The disk contains a gaseous and a solid component (comprising dust particles, pebbles, or boulders) and an embedded planet.
The solid component of the disk was {treated} as a pressureless fluid. 
The dynamics of {the} gaseous and solid components perturbed by an embedded planet are described by the following equations:

\begin{equation}
  \frac{\partial \Sigma_\mathrm{g}}{\partial t}+\nabla \cdot (\Sigma_\mathrm{g} {\textbf{v}})=0, 
  \label{eq:cont}
\end{equation}
\begin{equation}
  \frac{\partial \textbf{v}}{\partial t}+(\textbf{v} \cdot 
\nabla)\textbf{v}=-\frac{1}{\Sigma_\mathrm{g}} \nabla P +\nabla \cdot \textbf{T} -  \nabla \Phi-\frac{\Sigma_\mathrm{d}}{\Sigma_\mathrm{g}}\textbf{f}_\mathrm{drag},         \label{eq:NS}
\end{equation} 
\begin{equation}
  \frac{\partial \Sigma_\mathrm{d}}{\partial t}+\nabla \cdot 
    (\Sigma_\mathrm{d} \textbf{u})=- \dot{\Sigma}_\mathrm{acc} -\nabla\cdot\textbf{j}, 
    \label{eq:contd}
\end{equation}
\begin{equation}
   \frac{\partial \textbf{u}}{\partial t}+(\textbf{u} \cdot \nabla)\textbf{u}=- \nabla\Phi+\textbf{f}_\mathrm{drag}-(\textbf{u}\cdot\nabla)\textbf{j}, 
   \label{eq:NSd}
\end{equation} 
where $\Sigma_\mathrm{g}$, $\Sigma_\mathrm{d}$, and $\textbf{v}$, $\textbf{u}$ are the surface mass densities and velocity vectors of the gas and solid components, respectively. 
The disk was assumed to be locally isothermal{, in which} case the gas pressure was given by $P=c_\mathrm{s}^2\Sigma$, where $\mathrm{c_s}$ is the local sound speed.
A flat disk approximation was used{, in which} case the disk pressure {scale height} was defined as $H=hR$, where $h$, the disk aspect ratio, was set to 0.05.

The viscous stress tensor, $\textbf{T}$, in Eq.~(\ref{eq:NS}) was calculated as
\begin{equation}
    \textbf{T} = \nu\left(\nabla \textbf{v} + \nabla \textbf{v}^T -\frac{2}{3}(\nabla\cdot\textbf{v})\textbf{I}\right),
\end{equation}
where $\nu$ is the disk viscosity and $\textbf{I}$ is the two-dimensional unit tensor (see details in \cite{Masset2002} for {the calculation of} $\textbf{T}$ in {the} cylindrical coordinate system). 
We used the $\alpha$ prescription {from} \cite{ShakuraShunyaev1973} for the disk viscosity, in which case $\nu = \alpha c_s H$.
{We used} $\alpha =10^{-4}$, which is {on} the order of the smallest effective viscosity {thought to exist} in protoplanetary disks; arising, for example, due to {the} vertical shear instability \citep{StollKley2016}.

The gravitational potential of the system, $\Phi$, was calculated as
\begin{equation}
\Phi=-G\frac{M_*}{R}-G\frac{M_\mathrm{p}}{\sqrt{R^2+R_\mathrm{p}^2 - 2RR_\mathrm{p} \cos(\phi-\phi_\mathrm{p}) + (\epsilon H)^2}}+\Phi_\mathrm{ind},
\label{eq:phi_tot}
\end{equation} 
where $G$ is the gravitational constant, and $R$, $\phi$ and $R_\mathrm{p}$, $\phi_\mathrm{p}$ are the radial and azimuthal coordinates of a given numerical grid cell and the planet, respectively. 
$M_*$ and $M_\mathrm{p}$ {are the masses of} the star and planet, respectively. 
The indirect potential, $\Phi_\mathrm{ind}$, was taken into account, {since} a non-inertial frame corotating with the planet was used for the simulations. 
We note, however, that the effect of $\Phi_\mathrm{ind}$ {is expected to be negligible} in our case{, since} the {studied} planet-star mass ratio is small, $M_\mathrm{p}/M_*\leq10^{-5}$, and no significant large-scale asymmetries develop in the disk (see more details in  \citealp{RegalyVorobyov2017}). 

{The self-gravity of both the gas and the solid was} neglected. 
The planetary potential was smoothed by a factor of $\epsilon H$, assuming $\epsilon=0.6$, which {has been} found to be appropriate for two-dimensional simulations \citep{Kleyetal2012,Mulleretal2012}.

The {back-reaction} of {the} solid material is expressed by the last term in Eq.~(\ref{eq:NS}), {where} the drag force per unit mass between the dust and {the} gas is
\begin{equation}
    \textbf{f}_\mathrm{drag} = \frac{\textbf{v}-\textbf{u}}{\mathrm{St}/\Omega},
    \label{drag}
\end{equation}
where St is the Stokes number of the given solid species and $\Omega=(GM_*/R^3)^{1/2}$ is the local Keplerian angular velocity.
{To solve} Eq.~(\ref{eq:NSd}), we used a fully implicit scheme described in 
\citet{Stoyanovskayaetal2017,Stoyanovskayaetal2018} that was successfully tested in \citet{Pierensetal2019} and \citet{Vorobyovetal2019}.
First, the source term -- the right-hand side of Eq.~(\ref{eq:NSd}) -- was calculated, followed by the conventional advection calculation.
The solid velocity components were updated at {each} time step as
\begin{equation}
    \frac{\textbf{v}^{(n+1)}-\textbf{v}^{(n)}}{\Delta t}=\textbf{a}_\mathrm{g}^{(n)}-\xi\frac{\textbf{v}^{(n+1)}-\textbf{u}^{(n+1)}}{\tau_\mathrm{s}},
\end{equation}
\begin{equation}
    \frac{\textbf{u}^{(n+1)}-\textbf{u}^{(n)}}{\Delta t}=\textbf{a}_\mathrm{d}^{(n)}+\frac{\textbf{v}^{(n+1)}-\textbf{u}^{(n+1)}}{\tau_\mathrm{s}},
\end{equation}
where $\tau_\mathrm{s}=\mathrm{St}/\Omega$ is the stopping time of the given solid species,
$\Delta t$ is the {applied time step}, and $\textbf{a}_\mathrm{g}^{(n)}$ and $\textbf{a}_\mathrm{d}^{(n)}$ are the acceleration vectors of gas and solid {material} due to the pressure gradient and gravitational forces,
\begin{equation}
    \textbf{a}_\mathrm{g}^{(n)}=-\frac{1}{\Sigma_\mathrm{g}^{(n)}}\nabla P^{(n)}-\nabla \Phi^{(n)},
\end{equation}
\begin{equation}
    \textbf{a}_\mathrm{d}^{(n)}=-\nabla\Phi^{(n)}.
\end{equation}

The diffusion of solid material was taken into account; this was expressed by
\begin{equation}
    \textbf{j} = -D(\Sigma_\mathrm{g}+\Sigma_\mathrm{d})\nabla\frac{\Sigma_\mathrm{d}}{\Sigma_\mathrm{g}+\Sigma_\mathrm{d}},
\end{equation}
where \textbf{j} is the diffusion flux and 
\begin{equation}
    D=\nu/(1+\mathrm{St}^2)
    \label{diff}
\end{equation}
is the diffusion coefficient. 
In essence, the diffusion coefficient approximates the effect of the {small-scale} (much smaller than the size of the hydrodynamic grid) dynamics of the gas on the solid material. 
As a result, a local accumulation of {solid material} was smoothed out on the diffusion timescale ({which is} half the viscous timescale for $\mathrm{St}= 1$).

{The accretion of solid material onto the planet, represented by $\dot{\Sigma}_{\mathrm{acc}}$ in Eq.~(\ref{eq:contd}), was modeled by a reduction in the solid density {within} the planetary Hill sphere.
We used a scheme similar to the {gas accretion prescription} given in \citet{Kleyetal2012}.
The method used here for planetary accretion is the same as in \citet{Regaly2020}. The solid density was reduced by $1-\eta dt$ at each time step inside the Hill sphere of the planet with radius $R_\mathrm{Hill} = a(M_\mathrm{p}/3M_{*})^{1/3}$, where $\eta$ is the accretion strength. With the above constraint, the half-emptying time is $log(2)/\eta$; that is, about two thirds of the orbital time for $\eta= 1$. Two accretion scenarios were investigated, $\eta = 1$ and $0.1$, referred to as strong and weak accretion scenarios. The solid density reduction was done in two steps: first, one third of the density was removed from the inner part of the Hill sphere ($\Delta R < 0.75 R_\mathrm{Hill}$), then in the second step, two thirds of the density was removed from the innermost part of the Hill sphere ($\Delta R < 0.45R_\mathrm{Hill}$). As a result, the total solid mass and moment are strictly not conserved in the system.
The mass of solid material accreted by the planet is negligible compared to the planet mass for all solid species{, so} the accreted material was not added to the planet.
}

\subsection{Initial and boundary conditions}

We {studied} five different planetary masses in logarithmic bins $M_\mathrm{p} = 0.1,~0.3,~1,~3.3,$ and $10~M_\oplus$. 
The orbital distance of the planet was set to one in code {units}, and the planet was kept on a circular orbit; that is, no migration was allowed.

We {treated solids} with fixed Stokes numbers throughout the simulation domain. 
We modeled {several} solid species in eight bins: $\mathrm{St}= 0.01,~0.1,~1,~2,~3,~4,~5,$ and 10.
The reason for the increased resolution between $\mathrm{St}=1-5$ is that these solid species exhibit considerable variability in their solid torques.
We note that by adopting these values of the Stokes number, we covered the sizes belonging to {the} solid material from {the} millimeter to meter regime (see Fig~1. in \citealp{Regaly2020} for more details).

Initially, $\Sigma_\mathrm{g} = \Sigma_0 R^{-p}$ and $\Sigma_\mathrm{d} = \xi \Sigma_\mathrm{g}$ was set assuming $\Sigma_0 = 6.45\times10^{-4}$ and a {solid-to-gas} mass ratio of $\xi=10^{-2}$. 
We {explored} three different initial gas and solid density profile {slopes}: $p = 0.5,~1.0,$ and 1.5, with the above fixed value of $\Sigma_0$.

The initial velocity components of {the} gas ($v_R$ and $v_{\phi}$) were defined
as
\begin{equation}
    v_{R} = -3\alpha h^{2} (1-p) \Omega R,
    \label{eq:Vr}
\end{equation}
\begin{equation}
    v_{\phi} = \sqrt{1-h^{2} ( 1 + p )} \Omega R.
    \label{eq:Vphi}
\end{equation}
The above equations satisfy the steady-state solution {of} the viscous evolution of the surface mass density in $\alpha$ disks for $p = 0.5$ and $p = 1$. 
The initial velocity components of {the} solid material ($u_R$ and $u_{\phi}$ ) are given by the analytic solution of an unperturbed disk, which reads
\begin{equation}
    u_{R}= \frac{v_{R} \mathrm{St}^{-1}-h^{2}(1+p)}{\mathrm{St} + \mathrm{St}^{-1}} \Omega R,
\end{equation}
\begin{equation}
    u_{\phi} = \sqrt{1-h^{2}(1+p)} \Omega R-\frac{1}{2}u_{R}\mathrm{St},
\end{equation}
(see, e.g., \citealp{Nakagawaetal1986}; \citealp{TakeuchiLin2002}).

According to \citet{Garateetal2019}, if {the dust back-reaction is taken into account}, the initial velocity components of the gas given by (\ref{eq:Vr}) and (\ref{eq:Vphi}) should be modified as 
\begin{equation}
    v_{R}^*=A v_\mathrm{R} - 2B v_\phi,
\end{equation}
\begin{equation}
    v_{\phi}^*=\frac{1}{2} B v_R + A v_\mathrm{\phi}.
\end{equation}
Since in our simulations we assume that the disk contains only {one} solid {component} per simulation, the factors $A$ and $B$ are as follows:
\begin{equation}
    A=\frac{\xi+1+\mathrm{St}^2}{(\xi+1)^2+\mathrm{St}^2},
\end{equation}
\begin{equation}
    B=\frac{\xi\mathrm{St}}{(\xi+1)^2+\mathrm{St}^2},
\end{equation}
as was shown in \citet{Garateetal2020}.

Our models {use} an arithmetic grid with a numerical resolution of $N_R~\times~N_{\phi}$= $1536~\times~3072$ with a radial range of $[0.48-2.08]$.
{At} the above numerical resolution, the solutions are in the numerically convergent regime, as was tested earlier in \citet{Regaly2020}. 
{Thus, the planetary Hill sphere was resolved by about 1459, 314, and 68 cells for $10~M_\oplus$, $1~M_\oplus$, and $0.1~M_\oplus$ planets, respectively.}

At the inner and outer boundaries, a wave damping boundary condition was applied for the gas (see, e.g., \citealp{deValBorroetal2006}). 
For the solid material, we used an open boundary condition at the inner boundary.
Thus, the solid material is continuously lost to the star with the strongest rate for $\mathrm{St=1}$. 
{At the outer boundary, however, the dust replenishes} due to the applied damping boundary condition.  
As a result, the disk is not emptied by the solid material.

\subsection{Torque calculation}

The torques exerted on the planet by the solid and gas {components} were calculated:
\begin{equation}
    \Gamma =  \sum_{i,j=1}^{N_r,N_\phi} \Bigg( x_p \frac{\Sigma_{i,j}}{\Delta R^3_{i,j}} -y_p \frac{\Sigma_{i,j}}{\Delta R^3_{i,j}} \Bigg),
\end{equation}
where
\begin{equation}
    \Delta R_{i,j} = \sqrt{\Delta x^2_{i,j}+\Delta y^2_{i,j}},
\end{equation}
\begin{equation}
    \Delta x_{i,j} = R_{i,j} \cos(\phi_{i,j})-x_p,
\end{equation}
\begin{equation}
    \Delta y_{i,j} = R_{i,j} \sin(\phi_{i,j})-y_p,
\end{equation}
where $R_{i,j}$ and $\phi_{i,j}$ are the cylindrical coordinates of cell $i$, and $j$, $\Sigma_{i,j}$ is the surface density of {the} solid or gas {in} that cell, and $x_p$,$y_p$ are the Cartesian coordinates of the planet.
To compare {the} torques felt by {planets of different masses}, we normalized the torques.
The change {in a planet's} semi-major axis, $~a$, due to the torque, $\Gamma$, exerted by the gas and solid species can be given as $\dot{a}=2\Gamma/(qa\Omega)$, where $q$ is the planet-to-star mass ratio. Thus, $\Gamma_0 = 2/(qa\Omega)$ {will be} used as a normalization factor throughout this paper.
Since $a=1$ in our models, the normalization constant is $\Gamma_0=2/q$.

\section{Results} \label{chapt:3}

\subsection{Effect of back-reaction of solid material for an Earth-mass planet}

\begin{figure}
\includegraphics[width=9 cm]{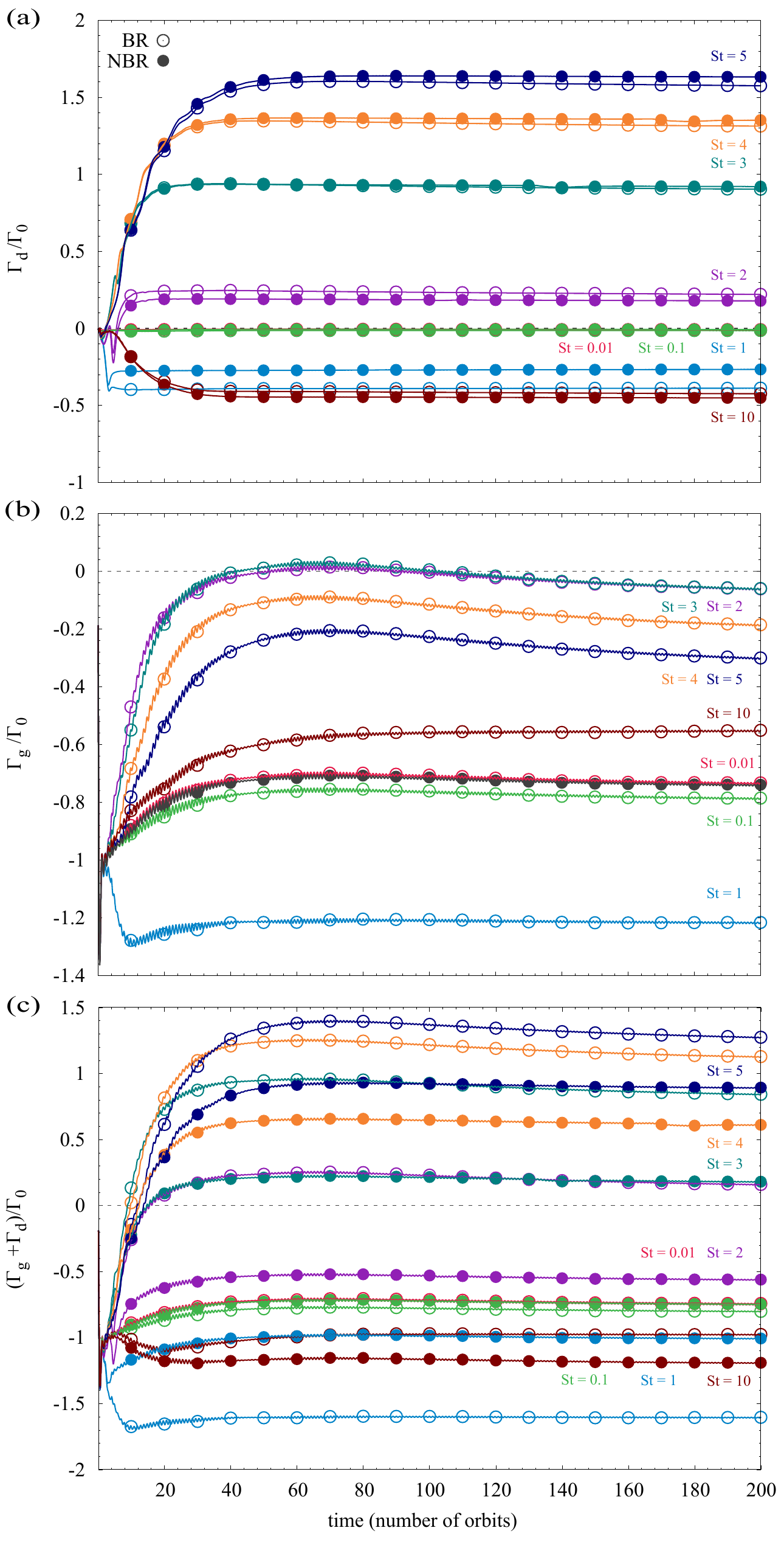}

\caption{Evolution of {the} solid (panel~a), gas (panel~b), and total {torques} (panel~c) exerted on a 1 $M_\oplus$ planet over time (measured in terms of the number of orbits). 
The torques are normalized by the factor $\Gamma_0=2/q$.
The different models -- with and without the back-reaction of {the} solid material -- are represented by the empty and filled circles, respectively. 
The colors indicate the different {types} of solid material.
}
\label{fig:3pic}
\end{figure}

First, we examined the effect of back-reaction on the torques felt by a $M_\mathrm{p}= 1~M_\oplus$ planet in a non-accreting model, assuming $p = 0.5$. 
Panel~a of Fig. \ref{fig:3pic} shows the evolution of {the} solid torques exerted by eight different solid species in non-back-reaction (NBR) and back-reaction (BR) models. 
As with the NBR models, the  {torque saturation} in {the} BR models also occurs within 200 orbits of the planet.
The time {required} for {the} solid torque {to saturate}  increases slightly with {the} Stokes number.
In the case of the coupled species ($\mathrm{St}= 0.01$ and 0.1), the {saturation value} of {the} solid torques {does} not change significantly in {the} BR models.
For $\mathrm{St}= 3,~4,$ and 5, less positive solid {torques} can be measured, while for $\mathrm{St}= 2$ the solid torque is more positive in BR models.
It is noteworthy that the shift {in} the saturation value of {the} solid torque resulting from the back-reaction is strongest for $\mathrm{St}= 1$.

Panel b of Fig. \ref{fig:3pic} shows the torques exerted by the gas interacting with eight different {types of solids in the} NBR and BR models.
The saturation of gas torques {also occurs} within 200 orbits of the planet.
Since {the} NBR model assumes that the dynamics of gas are independent of the solid material, the gas torques resulting from the simulations are the same for all nine solid species (black curve).
However, this is not the case for BR models, where the gas torque depends on the Stokes number of {the} solid material.
The shifts in the saturation value of {the} gas torques in {the} BR models are most {pronounced} for species with $\mathrm{St}= 1-5$.
For the coupled pebbles ($\mathrm{St}=0.01$), the change in gas torque is negligible compared to the NBR model.
However, for species with $\mathrm{St}= 0.1$, the gas torque is slightly more negative.
For $\mathrm{St}= 1$, the magnitude of the negative gas torque {increases significantly} (about 20 percent) in {the} BR models.
It is of particular interest that in the BR models {considering} species with $\mathrm{St}= 2-10$, the magnitude of the negative gas torques {is} significantly reduced as a consequence of {the} back-reaction.
The reduction in the magnitude of the negative gas torques is most pronounced for species with $\mathrm{St}= 2$ and 3, and then {decreases} with increasing Stokes number.

Next, we examined the total torque exerted on a 1 $M_\oplus$ planet, shown in panel c of Fig. \ref{fig:3pic}.
The influence of back-reaction {results in} a torque reversal for species with $\mathrm{St}= 2$, {compared} to the NBR model.
As a result of back-reaction, the magnitude of the positive total torques {is strengthened} for $\mathrm{St}= 3,~4,$ and 5.
For $\mathrm{St}=~10$, the total torque remains negative but its magnitude {reduced compared} to the NBR models.
Conversely, for $\mathrm{St}=~0.1$ and 1, the back-reaction of {the} solid material increases the magnitude of the negative total torque.
However, for coupled solid species, $\mathrm{St}=~0.01$, the total torques do not change due to the effect of back-reaction. 
In summary, the back-reaction of solid material significantly affects both the gas and {the} total torque experienced by an Earth-mass planet.
In the following, we extend our analysis to planets with different masses and different {solid material accretion efficiencies}.

\subsection{Parameter study on the effect of back-reaction}

\begin{figure*}
\includegraphics[width=18 cm]{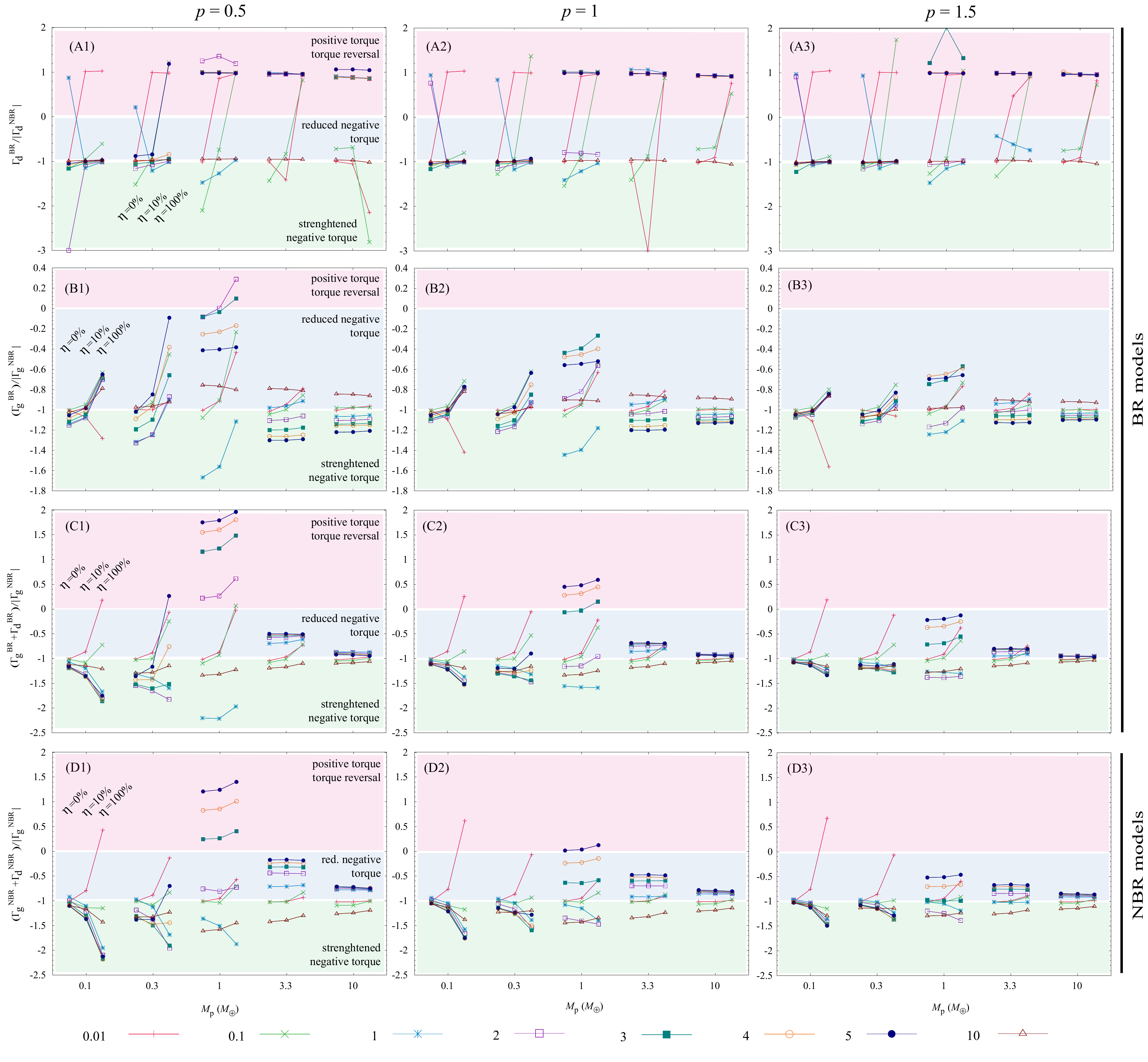}
\caption{Parameter study {of} the saturated torques felt by a low-mass planet in the range of  $0.1~M_{\oplus} \leq M_\mathrm{{p}} \leq 10~M_{\oplus}$.
{Panels A1,~A2,~A3:} solid torques in {the} BR models normalized by the absolute value of {the} solid torque in the corresponding NBR models.
{Panels B1,~B2,~B3:} gas torques in {the} BR models normalized by the {absolute value of the gas torque} in the NBR models.
{Panels C1,~C2,~C3:} total torques in {the} BR models normalized by the {absolute value of the gas torque} in the NBR models.
{Panels D1,~D2,~D3:} total torques in {the} NBR models normalized by the absolute value of the gas torque in the NBR models.
{The} columns from left to right show three {sets} of models, each assuming a different steepness of the initial disk density profile, $p = 0.5, 1.0,$ and 1.5.
For each planet mass, three accretion efficiencies are {examined}: $\eta= 0,~0.1$, and 1.
The colors indicate {the} different Stokes numbers ($\mathrm{St}=0.01,~0.1,~1,~2,~3,~4,~5,$ and 10) used for the solid species in the parameter study.}
\label{fig:tot_trq}
\end{figure*}

Figure \ref{fig:tot_trq} shows the normalized solid (A1,~A2,~A3), gas (B1,~B2,~B3), and total (C1,~C2,~C3) {torques} measured at the end of {the} simulations, after 200 orbits of the planet.
{Similar} to \cite{Regaly2020}, five different planet {masses} ($0.1,~0.3,~1,~3.3,$ and 10 $M_\oplus$) with three different {accretion efficiencies}, $\mathrm{\eta}= 0,~0.1,$ and 1{, were investigated.} 
Simulations were {performed with} a different steepness of the disk surface mass density profile, assuming $p= 0.5,~1.0,$ and 1.5. 
For comparison, the last row (D1,~D2,~D3) of Fig. \ref{fig:tot_trq} shows the total torques without the effect of back-reaction.
In all cases, except for the solid torques in panels A1,~A2,~and A3, the torques were normalized by the absolute value of the gas torque measured in the NBR models, $|\Gamma_\mathrm{g}^\mathrm{NBR}|$.
For the solid torques, the absolute value of the corresponding NBR solid torque, $|\Gamma_\mathrm{d}^\mathrm{NBR}|$, was {used} as the normalization factor.

Based on the value of the normalized torques in the NBR case, three different regimes can be identified: a) $\Gamma^\mathrm{NBR}/|\Gamma^\mathrm{NBR}|>0$, shaded with red; b) $-1<\Gamma^\mathrm{NBR}/|\Gamma^\mathrm{NBR}|$~<~0, shaded with blue; and c) $\Gamma^\mathrm{NBR}/|\Gamma^\mathrm{NBR}|<-1$, shaded with green colors.
Torque measurements in the green or blue regions indicate that the disk exerts an increased {or} decreased {amount} of negative torque, respectively. 
Conversely, planets {experience} a positive torque from the disk if the measured torques are in the red region.
Thus, at the boundary of the green and blue regions, the measured torques correspond to the canonical type I torques ($\Gamma / \Gamma_0=-1$). 

Panels A1, A2, and A3 of Fig. \ref{fig:tot_trq} {show} the {saturation} values of {the} solid torques in the BR scenario, measured at the end of {the} simulation.
In general, the back-reaction of {the} solid material has a marked effect on the solid torques for the coupled pebbles, $\mathrm{St}\leq0.1$, for all {the} planet {masses}, accretion {efficiencies}, and initial density {steepness} investigated.
Interestingly, {there can be a torque reversal} (where $\Gamma_\mathrm{d}^\mathrm{BR}/|\Gamma_\mathrm{d}^\mathrm{NBR}|<0$) for the coupled species due to accretion and back-reaction.
We also {note} that $M_\mathrm{p}=10~M_\oplus$ planets experience a significantly {enhanced} negative solid torque from the coupled species in BR models assuming $p=0.5$.
However, {there is} a weakened negative (for weak accretion, $\eta\leq0.1$) and positive (for strong accretion, $\eta=1$) solid torque assuming $p=1$ and 1.5.
For the weakly coupled species, $\mathrm{St}\geq1$, the solid torques are {unaffected} ($\Gamma_\mathrm{d}^\mathrm{BR}/|\Gamma_\mathrm{d}^\mathrm{NBR}|\simeq 1$) by the back-reaction of solid material felt by  $M_\mathrm{p}\geq1~M_\oplus$ planets.
{Conversely, for $M_\mathrm{p}\leq 0.3~M_\oplus$ planets, the back-reaction has a significant effect on the solid torques.}
{We note} that the effect of accretion efficiency can be clearly identified for the coupled species, $\mathrm{St}\leq0.1$, while {it is} suppressed for $\mathrm{St}\geq1$.
It is {noteworthy} that for Earth-mass planets the solid torque of {the} species with $\mathrm{St}= 2$ undergoes a {sign change} for $p=1$ and 1.5 due to the {back-reaction effect}.

Panels B1,~B2, and B3 of Fig. \ref{fig:tot_trq} show the gas torques for the three different {steepnesses of the disk's initial gas profile}.
Our results {show} that the effect of back-reaction significantly {changes} the gas torques for all planetary masses, accretion efficiencies, and initial density {steepnesses studied}. In BR models, the effect of accretion tends to weaken the magnitude of {the}  gas torques exerted on $M_\mathrm{p} \leq 1~M_\oplus$ planets (i.e., the torques are in the blue regions).
{We note} that the back-reaction of solid material has the most significant effect on the gas torques exerted on Earth-mass planets.
The strongest effect of accretion on the magnitude of the gas torque is found for $M_\mathrm{p}=0.3~M_\oplus$ planets.
In certain cases ($\mathrm{St}=~2$ and 3, with $p=0.5$), positive gas {torques} can be measured on $M_\mathrm{p}=1~M_\oplus$ as a consequence of the combined effect of back-reaction and high accretion efficiency.
For $M_\mathrm{p} \geq 3.3~M_\oplus$ planets, the {effect} of accretion on the gas torque {is diminished}. 
However, the effect of back-reaction tends to {increase} the magnitude of the negative gas torques for planets in this mass range.

The sum of the gas and solid torques -- the total torques -- are shown in panels C1,~C2, and C3 of Fig. \ref{fig:tot_trq}. 
{In general}, the back-reaction effect reduces the magnitude of the total negative torques exerted on $M_\mathrm{p}= 0.1 - 0.3~M_\oplus$ planets compared to the NBR case.
For Earth-mass planets, the influence of back-reaction {increases} the overall magnitude of the total torque when it is positive and {decreases} it when it is negative. 
However, this trend does not {hold for} species with $\mathrm{St}=1$; in this case, the negative torque {is enhanced by} back-reaction.
{We note} that for shallow density profile disks, $p=0.5$, a torque reversal occurs in BR models for $M_\mathrm{p}=0.3~M_\oplus$ planets accreting species of $\mathrm{St}= 5$ with high efficiency and for $M_\mathrm{p}= 1~M_\oplus$ planets interacting with $\mathrm{St}=2$ species {regardless} of accretion efficiency.
We also find torque reversals in BR models for $p=1$, $\mathrm{St}=4$ and 5, regardless of accretion efficiency and for $\mathrm{St}= 3$ when $\mathrm{\eta}= 1$.
For $p= 1.5$, we no longer find positive total torques in {the} BR models (except {for} one model where $St=0.01$ and $\eta=1$), but the magnitudes of {the} negative total torques are significantly reduced compared to the NBR case.
For larger planetary masses ($M_\mathrm{p} \geq 3.3~M_\oplus$), the influence of back-reaction {reduces} the magnitude of the total negative torques for almost all species ($\mathrm{St}<10$), regardless of accretion efficiency.
However, for $\mathrm{St}=10$, the magnitude of {the} negative total negative torque is found {to increase}.

\section{Discussion}

\subsection{Torque profiles in the vicinity of the planet}

Here, we {examine} in detail the {relationship} between the measured torques and the torque profiles, as well as the density distributions around an Earth-mass planet, limiting our discussion to {models with} shallow initial density {profiles} of $p=0.5$. 
However, the same analyses can be applied to simulations with $p=1$ and $p=1.5$.

{To explain the} results presented in Sect. \ref{chapt:3}, {we first} analyzed the radial torque profiles in the {near-planetary regions}.
The azimuthally averaged torque profiles of solid species (Fig. \ref{fig:dust_eta}) and gas (Fig. \ref{fig:gas_eta}) in the vicinity of an Earth-mass planet as well as a detailed discussion can be found in the appendices~\ref{seq:Tprofsolid}-\ref{seq:Tprofgas}.
Our general observation is that the shape of the torque profiles is very similar for BR and NBR models. 
Although there {is a little variation} in the profile, the integrated torque values can {be significantly different}.
It is {noteworthy} that the solid torques can only originate from the region of {about} $\pm 3 \Delta~R_\mathrm{Hill}$ of the planet. In contrast, for the gas, this region extends to {about} $\pm 15 \Delta~R_\mathrm{Hill}$, {regardless} of whether the back-reaction is considered.

\subsection{Distribution of gas and solid material around the planet}

{We shall now} compare the distribution of solid and gaseous material surrounding an Earth-mass planet in {the} BR and NBR models. 
The plots illustrating the difference in solid (Figs. \ref{fig:com_dust}-\ref{fig:com_dust2}) and gas (Figs. \ref{fig:com_gas1}-\ref{fig:com_gas2}) density distributions for the BR and NBR cases 
can be found in Appendices~\ref{seq:ddistsolid}-\ref{seq:ddistgas}.

\begin{figure*}
    \centering
    \includegraphics[width=0.9\textwidth]{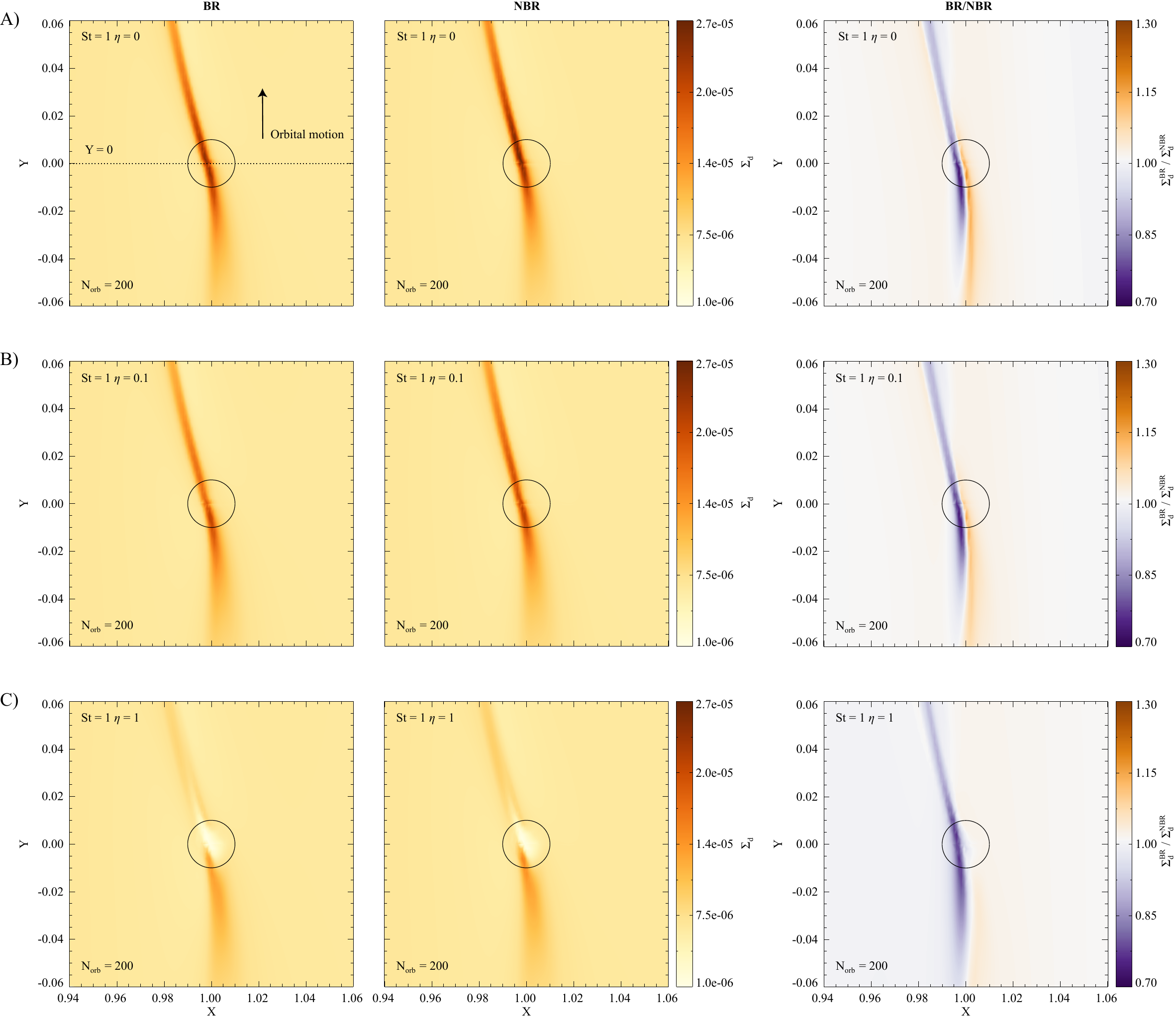}
    \caption{Distribution of $\mathrm{St}=1$ solid material in the vicinity of an $1~M_\oplus$ planet in BR and NBR models assuming $\eta=0$ (panels A), $\eta=0.1$ (panels B), and $\eta=1$ (panels C) with $p=0.5$. The density ratio of {the} BR and NBR models are also shown on the left. {The circles} represent the planetary Hill sphere. The axes are {shown} in code units.}
    \label{fig:dens-ratios}
\end{figure*}

The distribution of solid material {changes significantly} due to the combined effects of back-reaction and accretion (up to 60 percent density ratio for BR to NBR).
{An example is shown in Fig.~\ref{fig:dens-ratios} for an Earth-mass planet and $\mathrm{St}=1$ with $p=0.5$.
The effect of accretion as solid material becomes depleted close to the planet is clearly seen on the density distributions.
A comparison of the density ratios in the BR and NBR models ($\Sigma_\mathrm{d}^{\mathrm{BR}}/\Sigma_\mathrm{d}^{\mathrm{NBR}}$ shown in the left panels of Fig.~\ref{fig:dens-ratios}) shows that the effect of the back-reaction is to produce a density increase and decrease outside and inside the planetary orbit, respectively.
As is shown in Figs.~\ref{fig:com_dust} and \ref{fig:com_dust2}, a similar effect is observed for other Stokes numbers studied.
}
Conversely, the change in the distribution of gas due to the back-reaction of solid material and accretion {is less than 0.3 percent}.
Interestingly, the {change} in the distribution of solid material due to back-reaction does not {significantly affect} the magnitude of the solid torques (see Fig. \ref{fig:dust_eta}).
However, it is {clear} that even {small changes} in the {gas distribution} can lead to a {significant} reduction in the magnitude of the gas torque exerted on the planet.
To illustrate, the spiral waves generated by the gravitational influence {of the planet} are amplified in BR models for coupled pebbles with $\mathrm{St}\leq0.1$. 
In contrast, for pebbles or boulders with {stronger} coupling ($\mathrm{St}\geq1$), a {significant} depletion of gas is observed within the corotational region.
Furthermore, the strength of {the} accretion serves {to amplify the  above phenomena}.

\begin{figure*}
    \centering
    \includegraphics[width=0.9\linewidth]{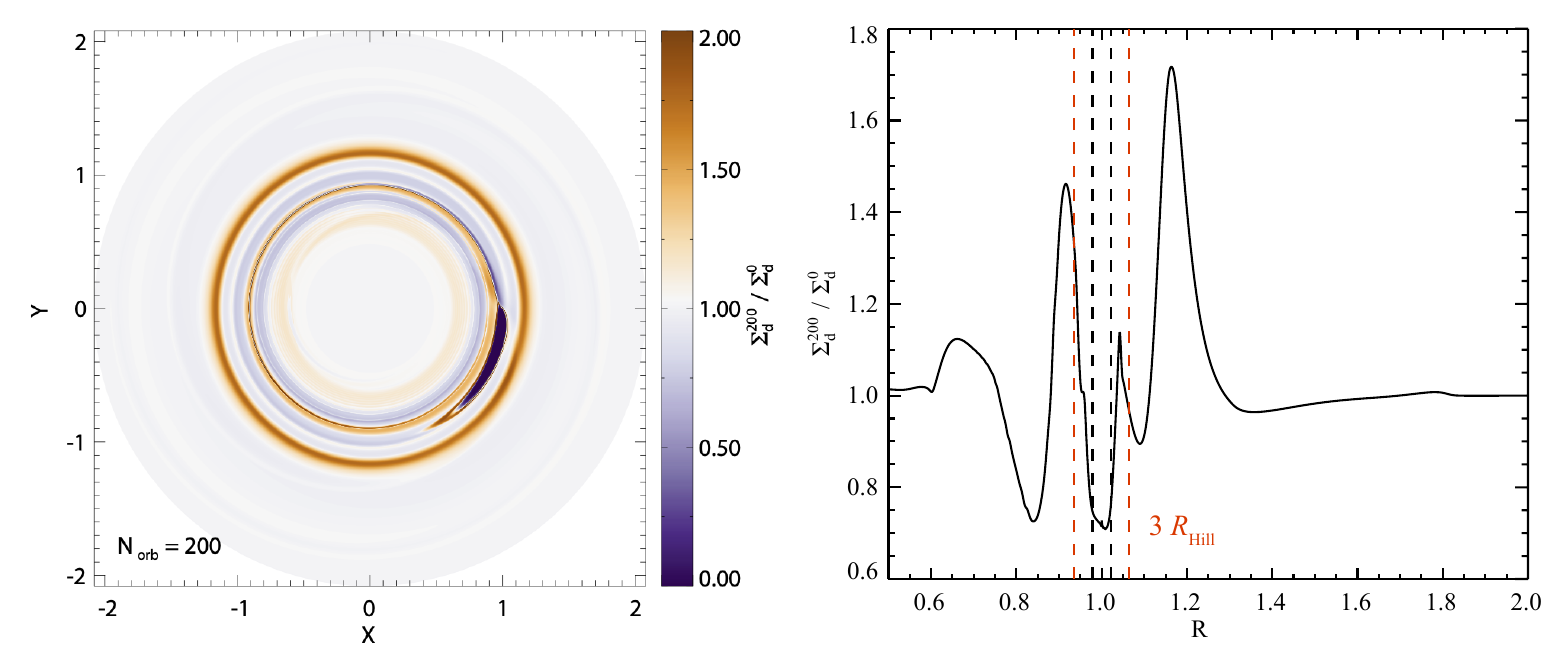}
    \caption{The normalized density distribution of the St=1 solid material after 200 orbits of a $10~M_\oplus$ planet, assuming $\eta=0$. Left: Two dimensional distribution. Right: Azimuthally averaged radial density profile of solid material. The inner and outer Hill radii are represented by dashed black lines.}
    \label{fig:isomass}
\end{figure*}

{Due to the low viscosity values used, planets with $10M_\oplus$ reach the isolation mass, as is shown in \citet{Ataiee2018A&A...615A.110A}. 
As a result, a weak (few-percent density depletion) partial gap is opened in the gas disk. 
The pressure maxima developed at the edges of the gap effectively collect solid material, especially for $\mathrm{St}=1$. 
An example of the solid material distribution as well as the azimuthally averaged radial density profiles for $\mathrm{St}=1$ and $\eta=0$ are shown in Fig.~\ref{fig:isomass}.
However, as we have shown (see e.g. Fig.~\ref{fig:dust_eta} for details), no significant torque is exerted by the solid material beyond the region $\Delta R > 3 R_\mathrm{Hill}$.
We note that in the long-term (several hundred thousand orbits) evolution of planet-disk interaction models, the corotation region can be emptied by the combined effect of solid material accretion and the pressure trap. 
As a result, the solid torque can be reduced, but it is already negligible for $10~M_\oplus$ planets.
}

\subsection{Caveats of our models}
\label{sec:Cav}

Finally, we discuss the caveats to our simulations that will be addressed in future research.
First, we modeled the planet-disk interaction in two dimensions for the case in which one has to smooth the gravitational potential of the planet defined by Eq.~(\ref{eq:phi_tot}).
The smoothing length was chosen to be $\epsilon=0.6$ {for both} gas and solid material.
We note that the vertical scale height of {the} solid material may differ from that of {the} gas due to {the} size-dependent {rate} of sedimentation to the disk midplane \citep{DullemondDominik2004}. 
However, {using} a planetary potential that depends on the Stokes number of solid material has no {firm basis} and requires further three-dimensional investigations. 
Thus, both solid material and gas experience the same smoothing of {the} planetary potential in our simulations.
\citet{RestrepoBarge2023} provided an analytical expression for the {spatially} varying smoothing length for two-dimensional protoplanetary disk simulations in which both dust and gas self-gravity are calculated.
As we showed in \citet{Regaly2020}, more or less smoothing reduces or increases the density asymmetry around the planet.
Recently, \citet{Chrenko2024A&A...690A..41C} {studied} the effect on the solid torques exerted by pebble-accreting low-mass planets, using Lagrangian superparticles to represent {the} pebbles.
In their model, therefore, no smoothing of the gravitational potential for pebbles is necessary.
Although they neglect the back-reaction of solid material, their results also show a positive total torque for $M_\mathrm{p}\leq3~M_\oplus$ planets for a range of Stokes numbers, $0.01\leq\mathrm{St}\leq0.785$.

We note, however, that three-dimensional models are necessary to model planetary accretion{, which is} a three-dimensional phenomenon in reality.
It has been shown that millimeter-sized dust in the circumplanetary region of a forming giant planet {does not settle in} the disk midplane  \citep{Szulagyietal2014,Fungetal2015,Szulagyietal2022,Cilibrasietal2023,Maedaetal2024}. 
As a result, {the} pebbles are delivered to the circumplanetary region {by} the meridional circulation accreted from the vertical direction, flowing together with the gas. 
Since this vertical mixing of the spiral waves can modify the two-dimensional angular momentum loss {three-dimensional} simulations should {also be considered for low-mass planets in the future}.
\citet{Fungetal2015} have also reported a migration rate of a $5~M_\oplus$ planet that is a factor of three slower due to three-dimensional effects, but this can be suppressed by a sufficiently large disk viscosity, leading to results similar to two-dimensional simulations.

{It is known that relatively large particles {with Stokes numbers} well above unity can cross each other's orbit. 
In this case, the fluid approach used in this study may neglect some aspects of orbital dynamics. 
In light of the above considerations, it would be preferable to use a comparable Lagrangian approach, as is used in \citep{Chrenko2024A&A...690A..41C}, instead of our current simplified methodology for modeling accretion phenomena. 
In this case, however, the diffusion of particles requires a particle-based implementation of diffusion, similar to the method used in \citet{Picognaetal2018A&A...616A.116P}. 
This will be left to a future study.}

The effect of thermodynamics is also neglected in our study {because} we used a locally isothermal approximation.
Since pebble accretion can transport significant energy to the planet, the heating of {the} circumplanetary material is inevitable.
As a result, a significant temperature contrast develops around the planet, {which can eventually cause} torque reversal \citep{PaardekooperMellema2006, Benitez2015}. 
{Two-dimensional} hydrodynamic simulations by \citet{KleyCrida2008} have already shown that low-mass planets ($M_\mathrm{p}\leq50~M_\oplus$) migrate outward in radiative disks {at} a rate comparable to that of standard type I migration.
This torque reversal {has also been} confirmed for planets in the range of $2-5~M_\oplus$  in three-dimensional hydrodynamic models \citep{Kleyetal2009,Legaetal2014}.
Thus, it is necessary to include the {thermodynamic} effects of solid accretion in future investigations.

We neglected the effect of the {disk's own gravity} by assuming that the disk mass is {small}.
However, in a more massive disk the results may differ from ours.
For example, \citet{Cridaetal2009} showed that neglecting the self-gravity of the disk requires careful handling of the material in the planet's Hill sphere. 
In our simulations, we did not apply {a torque cutoff} inside the Hill sphere.
This assumption is plausible, since the circumplanetary disk does not form around low-mass, $M_\mathrm{p}\lesssim 30~M_\oplus$, planets \citep{Massetetal2006}.
Thus, the exclusion of material in the planetary Hill sphere has no physical argument.
We emphasize that \citet{BaruteauMasset2008} showed that the disk self-gravity is slightly {accelerating} type I migration.
It should be noted that we have used arithmetically distributed radial grid cells, while the calculation of the self-gravity of the disk by fast Fourier algorithms (see, e.g., \citealp{VorobyovBasu2010,VorobyovBasu2015,RegalyVorobyov2017b}) requires logarithmically distributed radial grid cells. 

We utilized a {low-viscosity approach that deserves} further investigation at higher viscosities. 
The rationale behind this assertion is that in this particular scenario, the diffusion of solid material will be more pronounced, since diffusion is directly proportional to viscosity (Eq. \ref{diff}).
{As a result}, the asymmetry observed around the planet may be less pronounced, which {could} weaken the effect of the back-reaction of solid material. 
However, it should be noted that for larger Stokes numbers, diffusion will have a less significant effect, {since the diffusion coefficient} is inversely proportional to the square of {the} Stokes number. 

Furthermore, the analysis given in this paper {considers only} the accretion of solid material, while the accretion of gas is neglected.
{In addition}, the mass of solid material accreted by the planet is also not {included in} the planetary mass. 
However, {this  may} be a plausible assumption for low-mass planets, {since} we modeled the planet-disk interaction  for only 200 orbits, which limits the amount of material accreted {onto} the planet.

The low-mass planet was fixed in a circular orbit in our simulations; that is, no migration was allowed.
\citet{Guileraetal2023} calculated the formation tracks of low-mass embryos using the torque measurements of \citet{BenitezPessah2018}.
They found that low-mass planets ($\leq10~M_\oplus$) migrate outward beyond the water-ice line for $\mathrm{St}=0.1$ and provide the seed for massive planets.
It is therefore also worth investigating the effect of the back-reaction of the solid material combined with solid accretion in models in which planets are free to migrate.

Our models use a fixed Stokes number approximation for the solid species rather than a given size distribution for the solid material.
In the latter case, the Stokes number increases with stellar distance \citep{Weidenschilling1977}.
We note, however, that the radial range around the planet from which the solid torque is exerted is relatively small, only a few times the size of the planetary Hill sphere.
Nevertheless, it is desirable to calculate the local Stokes number based on the particle size when modeling evolutionary tracks.

Varying the metallicity (the dust content of the disk) can {change} our results, {since} the higher the metallicity, the higher the dust content of the disk.
For example \citet{HsiehLin2020} found stochastic migration of $2~M_\oplus$ planets for high metallicity ($Z>0.3$) and $\mathrm{St}\gtrsim0.03$; that is, for higher dust content than in our model.
It is therefore worth investigating the effect of disk metallicity on planetary torques in models that take into account the back-reaction of solid material.

\section{Conclusions}

We have studied how the back-reaction of solid material {affects} the solid, gas, and total torques exerted on low–mass ($0.1-10~M_\oplus$), solid-only accreting planets.
To achieve this, we performed two-dimensional hydrodynamic simulations that model the interaction {between the} accreting planet and the disk using the code {\small GFARGO2}.
We treated the solid material as a pressureless fluid and the disk was assumed to be locally isothermal.
The solid components of the disk were assumed to have fixed Stokes numbers in the range of $0.01 \leq \mathrm{St} \leq 10$.
{Diffusion of the} solid material was also taken into account. 
Simulations for different planetary masses and accretion efficiencies were performed with (BR models) and without (NBR models) the effect of the back-reaction of solid material.
The {saturation} value of {the torque} experienced by a non-migrating planet was determined {after} 200 orbits.
Our {findings} can be summarized as follows:
   \begin{enumerate}
      
      \item  In general, the magnitude of the solid torques is only slightly affected by the back-reaction.
      The {magnitude of the} solid torque  is observed to {decrease} in cases in which planets are accreting, with the potential for torque reversal to occur for those species that are still coupled to the gas, $\mathrm{St}\leq0.1$.
      These phenomena are independent of the {steepness} of disk density profile.
      
      \item In general, the gas torques are negative and {show} an increase in magnitude for all planetary masses {studied} under investigation in non-accreting models.
      It is noteworthy that Earth-mass planets {show} a distinctive behavior, with gas torques measured as being significantly weakened for $\mathrm{St}\geq2$.
      The {back-reaction effect increases} with {the} accretion efficiency and {decreases} with the steepness of {the} disk density profile.
      
      \item 
      In general, the magnitude of the total torque {is increased by} the effect of back-reaction, {regardless} of its sign.
      However, the combined effect of planetary accretion and back-reaction tends to {reduce} of the magnitude of negative torques.
     
      \item Compared to the canonical type I prediction, Mars-sized or slightly larger planets ($M_\mathrm{p} = 0.1-0.3~M_\oplus $) {with high accretion efficiency} experience a weakened magnitude torque or even a torque reversal for coupled pebble-sized solid material ($\mathrm{St}\leq0.1$).
      However, an {increased} magnitude of negative torque is measured for less coupled solid components ($\mathrm{St}\geq0.1$). 
      The steepness of the disk density profile has no significant effect on the above phenomena. 
      
      \item Earth-mass planets also experience a weakened magnitude {of negative torque} from the disk containing pebble-sized solid species ($\mathrm{St}\leq0.1$).
      However, the effect of back-reacting boulders ($\mathrm{St}=3-5$) is to strengthen the positive torque (or weaken the negative torque for steep disk density profiles), which requires a shallow disk density profile.
      {Conversely}, the negative torques are strengthened for $\mathrm{St}=1$ or $\mathrm{St}=10$.
      
      \item Super Earths ($M_\mathrm{p}\geq3.3~M_\oplus$) never experience positive total torques.
      The magnitudes of {the} total torques are still weakened compared to the type I prediction, however, {but} the effect of the back-reaction of {the} solid material {mitigates} the discrepancy.
      The accretion efficiency has no significant effect on the total torque.
      
   \end{enumerate}

As is {shown}, the back-reaction of {the} solid material has the effect of reducing the magnitude of the negative torques exerted on the planets, while {at the same time} increasing the magnitude of the positive torques exerted on the planets.
This phenomenon is most pronounced in {the case of} Earth-mass planets.
{Assuming} that our torque measurements are applicable to planets whose orbits are not fixed and that the pressureless fluid approximation of solid material is valid, the following train of thought can be {advanced}.

Mars-sized planetary embryos {that accrete} pebbles with high efficiency can migrate outward during their nascent phase when the solid material is still coupled to the gas of the protoplanetary disk.
Should the size of the solid components reach {the size of} boulders {during} this initial stage of planet formation, the planetary embryos {will undergo} a relatively {rapid} inward migration.
We note, however, that the {rate of} migration of these small embryos {at this stage of evolution is quite} slow.
Planets that {reach} a mass comparable to that of Earth are subject to outward migration due to the influence of solid material in the boulder regime.
Once a planet's mass exceeds that of the Earth, the direction of migration {is reversed} again, with a reduced rate compared to that predicted by the type I regime.

It should be emphasized that the deviation from the migration history predicted by the canonical type I theory may be more pronounced in disks with a higher metal content.
Furthermore, it is predicted that the negative total torques, which {have been} reduced in magnitude as a result of the back-reaction of solid material, can be further weakened and may become positive if the metal content of the disk is sufficiently high.
In light of the {above} findings, we conclude that planetary system formation {may} benefit from the combined effect of back-reaction and accretion by reducing the risk of young planets being engulfed by the central star. 

\begin{acknowledgements}
ZsS acknowledges the support of the NKFIH excellence grant TKP2021-NKTA-64.
\end{acknowledgements}

\bibliographystyle{aa}
\bibliography{ref}

\appendix

\section{Solid torque profiles for a $1~M_\oplus$ planet}
\label{seq:Tprofsolid}

\begin{figure*}[!h]
\centering
\includegraphics[width=2 \columnwidth]{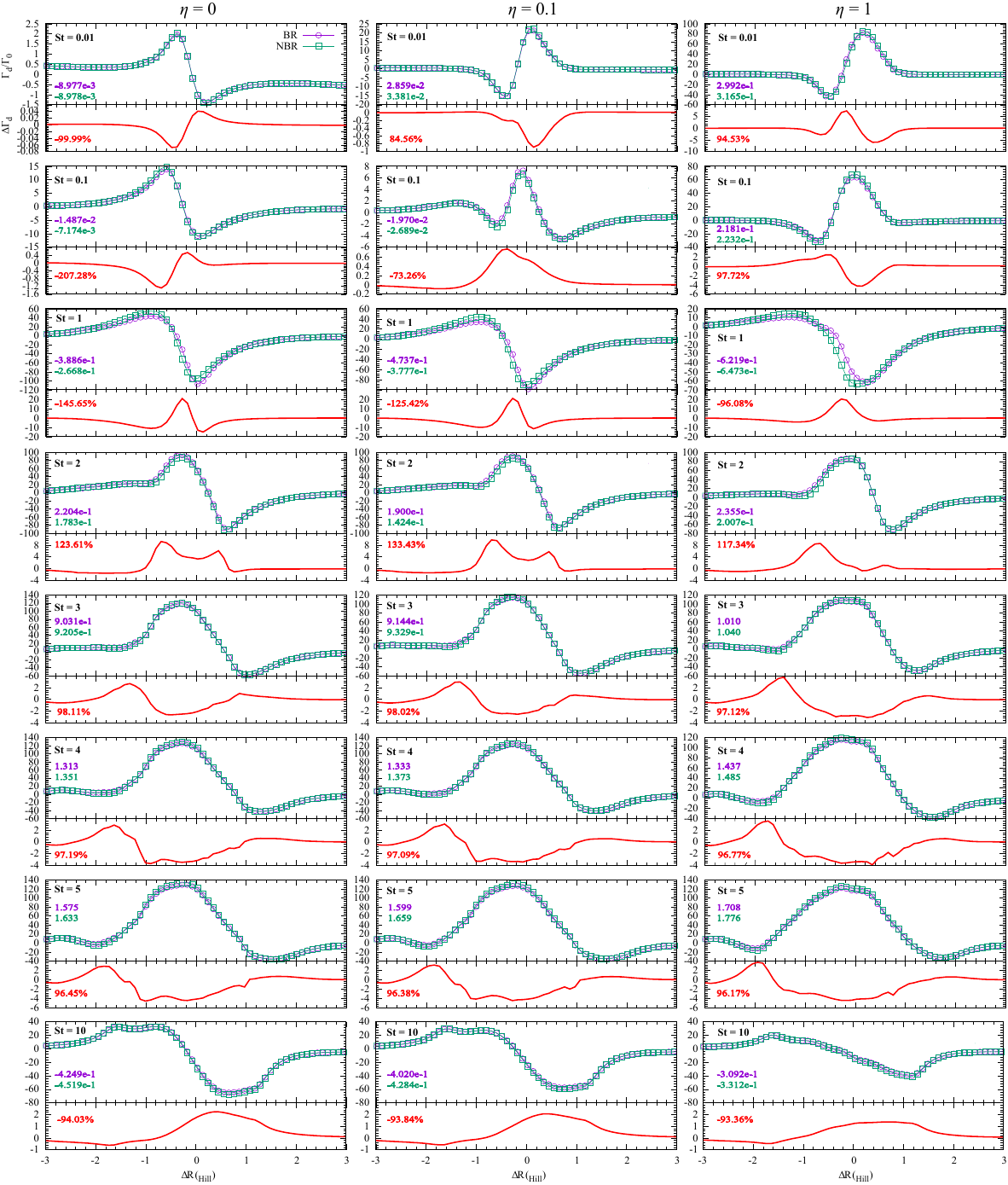}
\caption{Comparison of solid torque profiles for 1 $M_\oplus$ planet, assuming $p= 0.5$ and $\mathrm{\eta}= 0,~0.1$ and 1 in BR and NBR models. Note that the values {shown} on the vertical axis are multiplied by a factor of $10^{3}$ for the sake of clarity.}
\label{fig:dust_eta}
\end{figure*}

This section presents a detailed analysis of the solid and gas torque profiles in the vicinity of an Earth-mass planet, assuming $p= 0.5$ and $\mathrm{\eta}= 0,~0.1$ and 1.
The azimuthally averaged radial torque profile, $\Gamma_i$, is calculated as
\begin{equation}
    \Gamma_i = \frac{1}{N_{\phi}} \sum_{j=1}^{N_\phi} \frac{\Gamma_{i,j}}{\Gamma_0},
\end{equation}
where $\Gamma_{i,j} / \Gamma_0$ is the torque originating from cell $i,j$.
Figure \ref{fig:dust_eta} {shows} the azimuthally averaged solid torque profiles exerted by eight {different types of} of solids.
{The colors} indicate the two {different types of models}: those with (purple) and those without the back-reaction of solid material (green), respectively.
The difference in the measured torques between the two models, $\Delta\Gamma_\mathrm{d} = \Gamma_\mathrm{d}^\mathrm{NBR}-\Gamma_\mathrm{d}^\mathrm{NBR}$, is shown in red bellow each panel.
The sum of the BR and NBR torques along the X-axis is {shown} in the figures as purple and green numbers, respectively.
The model-dependent solid torque ratio in {percent} ({hereafter} referred to as solid torque ratio henceforth), $\Gamma_\mathrm{d}^\mathrm{BR}/|\Gamma_\mathrm{d}^\mathrm{NBR}|$, is {shown in} red  on each panel.
Note, that the {values displayed} on the vertical axis are multiplied by a factor of $10^{3}$ for the sake of clarity.
The horizontal axis {shows} the relative distance from the planet, expressed in units of the planet's Hill radius.
It is also important to note, that in the case of solid torque profiles, we only {show} the region close to the planet ($\pm 3~\Delta R_{\mathrm{Hill}}$).
This is because the solid torque is only significant up to this distance and generally {disappears} beyond $\pm 6~\Delta R_{\mathrm{Hill}}$ \citep{Regaly2020}.

First, we examined the torque profiles of {the} solid material assuming $\mathrm{\eta}= 0$ and $p=0.5$ (left column of Fig. \ref{fig:dust_eta}).
The shape of the solid torque profiles are the same for the NBR and BR cases for all {investigated species}.
The torques of the coupled solid species ($\mathrm{St} \leq 0.1$) {show small} discrepancies (less than 1 percent) in the torque amplitude between the BR and NBR scenarios.
While the change in the torque {amplitude} can reach up to about 207 percent for species with $\mathrm{St} = 0.1$, the solid torque {amplitudes} for the coupled species are so small (less than 1e-2), that this difference will not have a meaningful effect on {the} solid torque exerted on the planet.
This explains why the solid torques seen in our models (panel a of Fig. \ref{fig:3pic}) are the same for these species.

For $\mathrm{St}= 1$, the torque profile is shifted downward when {the} back-reaction is taken into account and the solid torque is about 146 percent that of observed in the NBR case.
As a result, the planet experiences an increase in the magnitude of {the} negative torque of this type (light blue curve in panel~b of Fig. \ref{fig:3pic}).

For $\mathrm{St}= 2$, {the} back-reaction causes the left wing of the positive peak and the negative peak of the solid torque to shift {upward}, {resulting} in a torque ratio of about 124 percent.
{As a result}, the planet experiences an increase in the magnitude of the positive solid torque in the BR case (purple curve in panel~a of Fig. \ref{fig:3pic}).

For species with $\mathrm{St}= 3-5$, the torque amplitude decreases in the {range}  [$-1\Delta R_{\mathrm{Hill}},+1\Delta R_{\mathrm{Hill}}$], while {it} increases outside {this range}.
For these weakly coupled species, the influence of {the} back-reaction is less pronounced, the torque ratio is only about 96-98 percent.
This explains the slightly {lower} saturation values for the BR torques of these species seen in panel a of Fig. \ref{fig:3pic}.
For species with $\mathrm{St}= 10$, the change in torque magnitude is the {smallest, about} $94$ percent of that observed in the NBR case (dark red curve on panel a of Fig. \ref{fig:3pic}).

Next we {examined} the effect of accretion on the solid torque profiles (middle and right columns of Fig. \ref{fig:dust_eta}).
In general, the effect of accretion has no {significant} impact on the shape of {the} radial solid torque profiles.

Weak accretion ($\mathrm{\eta}=0.1$) of species with $\mathrm{St}\leq0.1$ increases the torque amplitude inside the Hill sphere in the BR case {if} the sign of the torque is positive ($\mathrm{St}=0.01$) and decreases {if it is} negative ($\mathrm{St}=0.1$).
As a result both the positive and negative solid torques are weakened compared to the NBR case.
Strong accretion ($\eta =1$) {of} these species {reduces} the magnitude of the solid torques in the BR case.

For species with $\mathrm{St}= 1$, the difference in the torque magnitude decreases as the accretion efficiency increases.
This results in {a} the decrease {in} the solid torque ratio to about 125 and 96 percent {for} $\eta=0.1$ and 1, respectively.

For species with $\mathrm{St}= 2$, the torque magnitude is observed to increase to 133 percent of that in the NBR case when $\mathrm{\eta}= 0.1$, but only up to about 117 percent when the planet is {heavily} accreting ($\mathrm{\eta}= 1$) this solid material.
For species with $\mathrm{St}= 3-10$, the change in the torque ratio is negligible, {showing} a variation of only about 1 percent between the two different accretion scenarios.

\section{Gas torque profiles for a $1~M_\oplus$ planet}
\label{seq:Tprofgas}

\begin{figure*}[!h]
\includegraphics[width=2 \columnwidth]{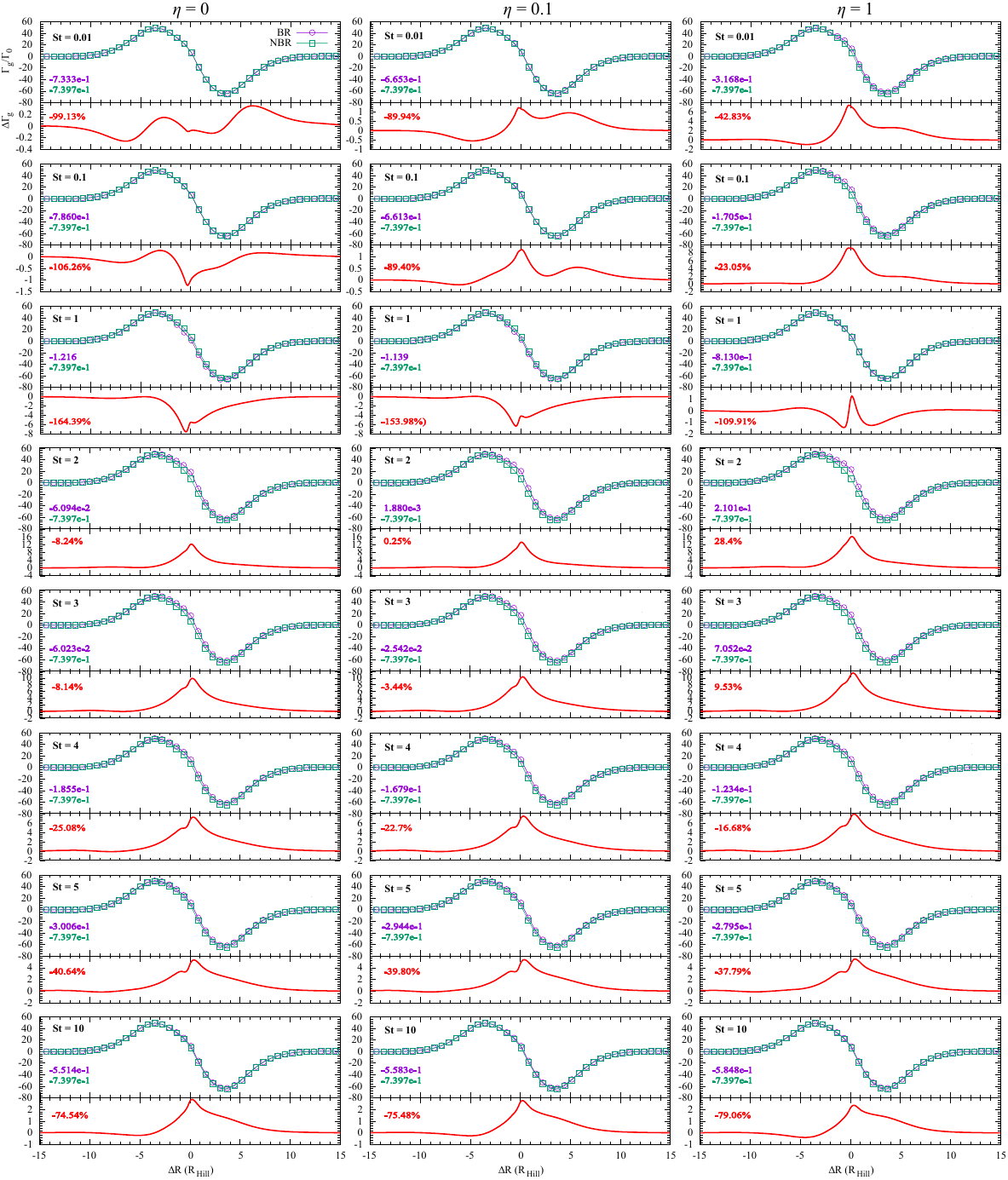}
\centering
\caption{Comparison of gas torque profiles for 1 $M_\oplus$ planet, assuming $p= 0.5$ and $\mathrm{\eta}= 0,~0.1$ and 1. in BR and NBR models. Note, that the {shown} values on the vertical axis are multiplied by a factor of $10^{3}$ for the sake of clarity.}
\label{fig:gas_eta}
\end{figure*}

Figure \ref{fig:gas_eta} shows the azimuthally averaged radial gas torque profiles caused by the interaction with different species ($\mathrm{St}= 0.01,~0.1,~1,~2,~3,~4,~5,$ and 10) of solid material in the vicinity of 1 $M_\oplus$ planet, assuming $p= 0.5$ and $\mathrm{\eta}= 0,~0.1,$ and 1.
Compared to the solid material, the radial torque of the gaseous component can {reach} or even {exceed}, $\pm 10~\Delta R_{\mathrm{Hill}}$ \citep{Regaly2020}. 
Therefore, a larger area must be displayed in order to study its variation ($\pm 15~\Delta R_{\mathrm{Hill}}$ in our case).
The colors indicate the two types of models, with (purple) and without {(green)} the back-reaction of {the} solid material, respectively.
The difference {between} the two models, $\Delta\Gamma_\mathrm{g} = \Gamma_\mathrm{g}^\mathrm{BR}-\Gamma_\mathrm{g}^\mathrm{NBR}$, is shown in red {below} each panel.
The {total} BR and NBR torques along the X-axis {and} the sum of their difference {are shown as purple, green and red} numbers, respectively.
The model-dependent ratio of {the} torques in percent ({hereafter} referred to as {the} gas torque ratio), $\Gamma_\mathrm{g}^\mathrm{BR}/|\Gamma_\mathrm{g}^\mathrm{NBR}|$, is represented by the red numbers in each panel.
Note that, as in the previous case, the values on the vertical axis are multiplied by $10^{3}$ for the sake of clarity.
The relative distance from the planet is {given} in units of the planet's Hill radius.

Next we {examined} the torque profiles assuming $\mathrm{\eta}= 0$ and $p= 0.5$ (left column of Fig. \ref{fig:gas_eta}).
Similar to solids, the shape of the gas torque profiles {is} also independent of whether the BR or NBR case is being considered.
For species with $\mathrm{St} \leq 1$, the sign of the gas torques remains negative and the gas torque ratio increases with the Stokes number, reaching a value of about 164 percent.

The back-reaction of species with $\mathrm{St}=1$ causes the gas torque amplitude to {decrease} in the {range} [$-3\Delta R_{\mathrm{Hill}},+8\Delta R_{\mathrm{Hill}}$].
This explains the increase in the saturation value of {the} negative gas torque shown in panel b of Fig. \ref{fig:3pic} for this solid species.
However, for the weakly coupled species ($\mathrm{St}= 2-10$), the gas torque amplitude is {increased} between $\pm 5~\Delta R_{\mathrm{Hill}}$ in the BR case. 
As a result, the gas torque ratio decreases significantly, 
{and reaches} a minimum value of about 8 percent that in the NBR case, when $\mathrm{St}= 2$ and 3.
This explains the significantly reduced {negative} saturation values in the BR case for these species (purple, green, yellow, blue and brown curves of panel b in Fig. \ref{fig:3pic}).
{In contrast} to the solid torque profiles, the change in the torque magnitude for these species decreases with Stokes number.

The middle and right columns of Fig. \ref{fig:gas_eta} {show} the azimuthally averaged radial gas torque profiles in two different accretion regimes ($\mathrm{\eta}= 0.1$ and 1, respectively).
It is {noteworthy} that in the case of the coupled species, solid accretion {leads to} a reduction in the magnitude of the negative gas torque.
For $\mathrm{St}= 0.01$ and 0.1, the torque magnitude in the weakly accreting scenario ($\mathrm{\eta}= 0.1$) is about 89 percent that of the NBR case.
Furthermore, in the strongly accreting scenario ($\mathrm{\eta}= 1$), the gas torque ratio {is significantly reduced by} the accretion of these species, down to about 42 and 23 percent, respectively.
For species with $\mathrm{St}= 1$, the torque ratio between BR and NBR cases decreases as the efficiency of solid accretion increases. 

In contrast to solids, the amplitude of the torque exerted by the gaseous component increases between $\pm 5~R_{\mathrm{Hill}}$ in the weakly and strongly accreting {scenarios}, when interacting with species {with} $\mathrm{St}= 2-10$.
As a result, the magnitude of the gas torque for these species continues to {decrease} compared to the NBR case.

For species with $\mathrm{St}= 2$, the magnitude of the negative gas torque is further reduced by accretion and undergoes a change in sign, reaching a magnitude of about 0.25 percent that of the NBR case when $\mathrm{\eta}= 0.1$.

For $\mathrm{St}= 3$, the gas torque is also positive in the BR case when $\mathrm{\eta}= 1$, with a gas torque ratio of about 10 percent.
It is significant {that} the gas torque becomes positive {in} these cases (see {the} purple and dark green symbols in panel B1 of Fig. \ref{fig:tot_trq}).
For species with $\mathrm{St}= 4$ and 5, the influence of back-reaction and accretion does not {lead to} torque reversal, {but} the negative torques are significantly reduced compared to the NBR case.
Similar to the non-accreting scenario ($\mathrm{\eta}= 0$), the change in the gas torque magnitude (i.e., the gas torque ratio) is observed to decrease with Stokes number for these species.
For species with $\mathrm{St}= 10$, the gas torque ratio increases slightly with the {accretion efficiency} to about 79 percent when $\mathrm{\eta}= 1$.

\section{Spatial distribution of solid material around an Earth-mass planet}
\label{seq:ddistsolid}

\begin{figure*}
\includegraphics[width=17 cm]{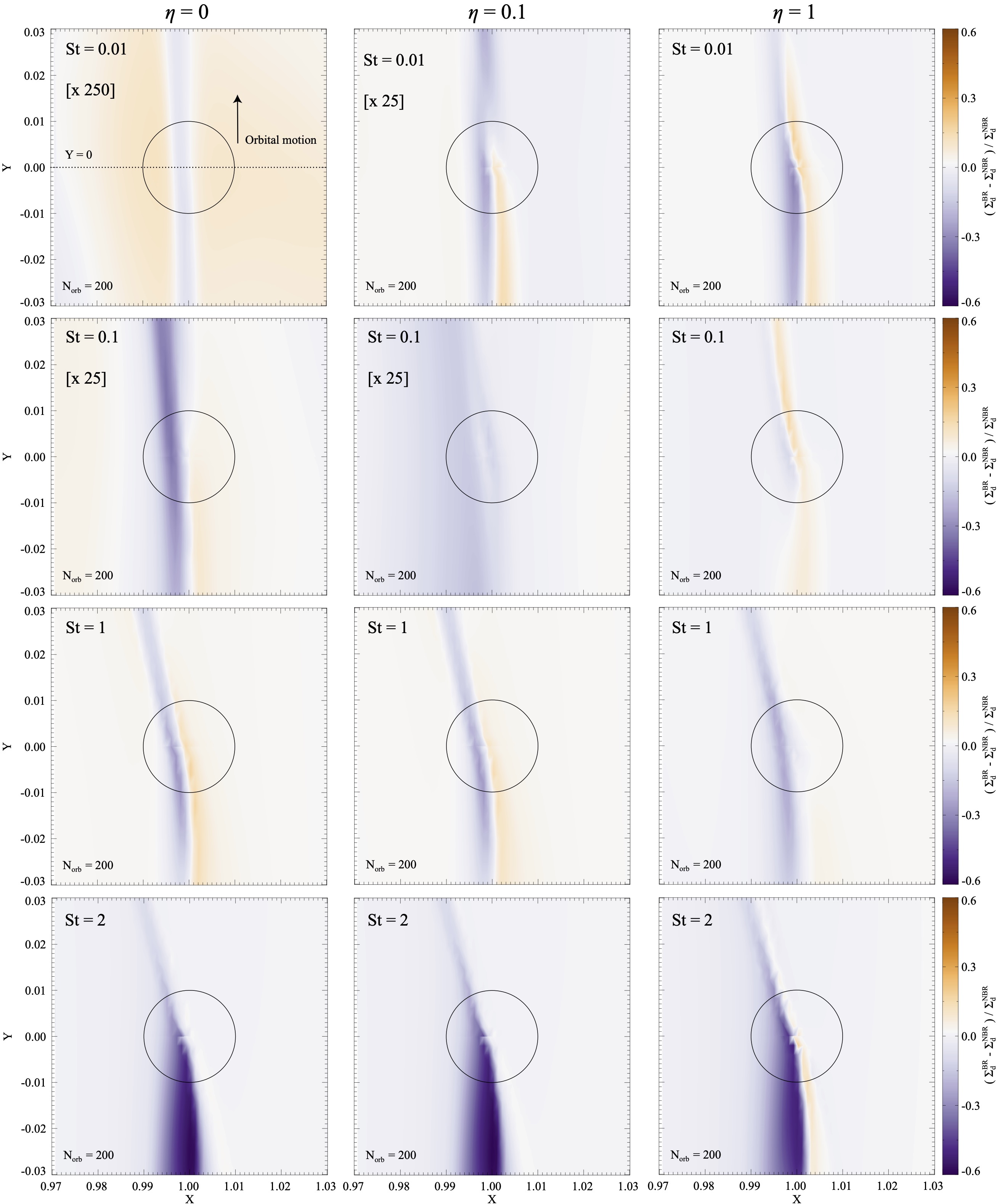}
\centering
\caption{Comparison of solid density distributions in NBR and BR models for $\mathrm{St}= 0.01,~0.1,~1,$ and 2, with three different accretion efficiencies ($\mathrm{\eta}= 0,~0.1,$ and 1), in the vicinity of 1 $M_\oplus$ planet.
The axes are {shown} in code units.
{Colors} ranging from dark purple to brown represent increasing solid densities in {the} BR models. Regions where the two models are the same {are shown} in white. The black circle indicates the planetary Hill sphere.}
\label{fig:com_dust}
\end{figure*}

\begin{figure*}
\includegraphics[width=17 cm]{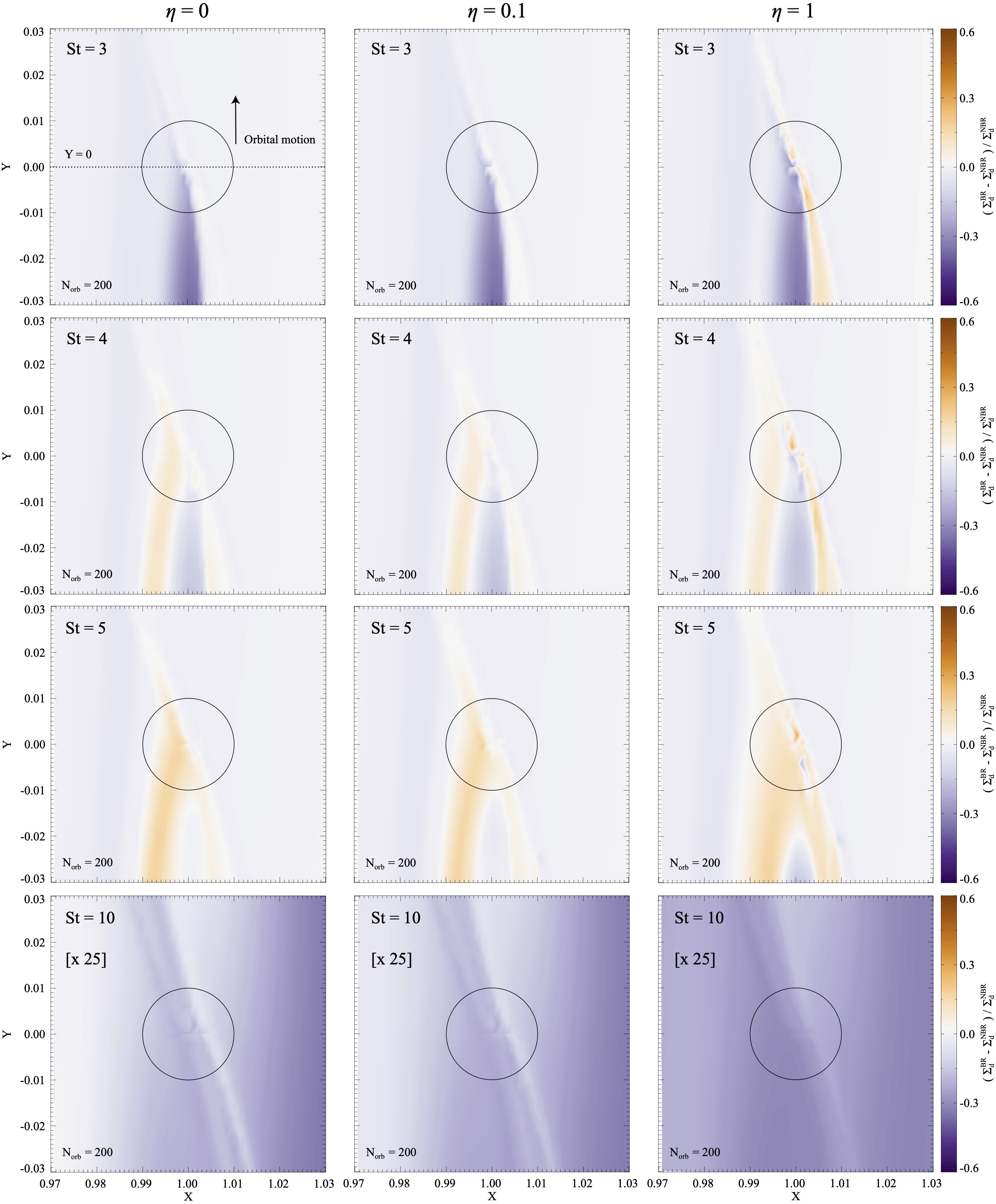}
\centering

\caption{Same as Fig.~\ref{fig:com_dust} but for $\mathrm{St}= 3,~4,~5,$ and 10.}
\label{fig:com_dust2}

\end{figure*}

{Figures} \ref{fig:com_dust} and \ref{fig:com_dust2} {show} the ratio of {the NBR to BR} solid surface density distribution in the vicinity of a $1M_\oplus$ planet, assuming $p=0.5$ and three different {accretion strengths}.
The density distributions are shown in a co-moving frame of the planet.
The material situated in the region {in front} ($Y>0$) or {behind} ($Y<0$) the planet (with respect to the orbital motion) exerts {a positive or negative torque}, respectively.
The rows represent the different {types} of solid species {studied} ($\mathrm{St}= 0.01,~0.1,~1,~2,~3,~4,~5,$ and 10), while the columns indicate increasing accretion efficiency ($\mathrm{\eta}= 0,~0.1,$ and 1). 
The comparison is calculated as $(\Sigma_\mathrm{d}^\mathrm{BR}-\Sigma_\mathrm{d}^\mathrm{NBR})/\Sigma_\mathrm{d}^\mathrm{NBR}$.
Note that {the} axis values are displayed in code units.

{We first} considered models {with} $\eta = 0$ (left column of Fig.~\ref{fig:com_dust}).
{Due to the remarkably} low comparison values observed {for} species with $\mathrm{St}= 0.01,$ and 0.1, the values {shown} in the figures have been multiplied by $250$ and 25, respectively. 
For $\mathrm{St}= 0.01$, even after {multiplying the} values, the change in the distribution between the BR and BR cases is very small.
Since this change is symmetrical to the $Y =0$ axis, the torques remain {unchanged} between the BR and NBR cases ($\mathrm{St}= 0.01$, $\eta = 0$ panel of Fig. \ref{fig:dust_eta}).
For $\mathrm{St}= 0.1$, the density of {the} solid material is {reduced in} the inner part of the corotation region in {the} BR model.
Conversely, a slight increase in the density of solid material is observed at the outer side of {the} corotation region, behind the planet.
Consequently, the distribution of solid material along the $Y =0$ axis will be asymmetric, resulting in a {BR to NBR} torque ratio of about 207 percent ($\mathrm{St}= 0.1$, $\eta = 0$ panel of Fig. \ref{fig:dust_eta}).
This phenomenon is more pronounced for species with $\mathrm{St}= 1$, which explains the measured torque ratio of about 146 percent and a negative torque that is increased in magnitude in the BR case ($\mathrm{St}= 1$, $\eta = 0$ panel of Fig. \ref{fig:dust_eta}).

For $\mathrm{St}= 2$, the disk is significantly reduced in solid material in the region behind the planet in the BR model.
This asymmetry along the $Y =0$ {explains} an explanation for the observed change in the torque ratio and the increase in the magnitude of the solid torque in the BR case.

For species with $\mathrm{St}= 3$, the region behind the planet in solid material {is still reduced} compared to the NBR case, {but this reduction} is less pronounced than in the case of $\mathrm{St}= 2$ (Fig.~\ref{fig:com_dust2}).
For species with $\mathrm{St}= 4$ and 5, the {empty} region behind the planet starts to {decrease} in {the} BR model.
Furthermore, in the region inside the planetary orbit, {there is} an enhancement {of} solid material develops in the BR case.
This results in an {increased} contribution of negative solid torque from the inner disk, which ultimately leads to a decrease in the magnitude of the solid torque compared to the NBR case.
The distribution of solid material becomes increasingly asymmetric with respect to the $Y =0$ axis as the Stokes number increases. 
This phenomenon is reflected in the weakening of the solid torque magnitudes for these species ($\mathrm{St}= 3-5$, $\eta = 0$ panel of Fig. \ref{fig:dust_eta}).

However, for species with $\mathrm{St}= 10$, there is a slight decrease in the density of {the} solid material behind the planet and outside the planetary orbit in the BR case. 
Note, that here the density values are multiplied by 25 for clarity.
Consequently, the magnitude of the negative torque decreases in the BR case ($\mathrm{St}= 10$, $\eta = 0$ panel of Fig. \ref{fig:dust_eta}).

\cite{Regaly2020} showed that the accretion of solid material {changes its} distribution around the planet and thus the magnitude of the solid torque exerted on the planet. 
Let's now compare the distribution of solid material in the BR and NBR cases in two accretion regimes, $\mathrm{\eta}=~0.1$ and 1 (middle and right columns of Figs. \ref{fig:com_dust}-\ref{fig:com_dust2}).
{Note} that for species with $\mathrm{St}= 0.01$ and 0.1, the density values on the figures are multiplied by 25 for $\eta = 0.1$.
As the accretion efficiency increases the density of solid material inside the corotation region {decreases slightly}, while {outside it increases} for $\mathrm{St}= 0.01$.
This increasing asymmetry in the solid material explains the measured 84 percent torque ratio between the BR and NBR cases.
The largest asymmetry {is} seen for $\eta = 1$.
For $\mathrm{St}= 0.1$, the combined effect of back-reaction and high accretion efficiency ($\mathrm{\eta}= 1$) results in an {increased} density of solid material in the corotation region in {the} BR models.
{However}, the aforementioned discrepancy will only result in a torque ratio of about 98 percent, {provided} that the change in the distribution of solid material remains symmetric {with respect} to the $Y =0$ axis of the planet ($\mathrm{St}= 0.1$, $\eta = 1$ panel of Fig. \ref{fig:dust_eta}).

For $\mathrm{St}= 1$, as the accretion efficiency increases, the density asymmetry of {the} sold material at the inner and outer corotation {regions} weakens. 
Consequently, the change in {the} distribution of solid material becomes increasingly {symmetric} with respect to the $Y =0$ axis, {resulting in} a reduction {of} the torque ratio to about 96 percent when $\mathrm{\eta}= 1$ ($\mathrm{St}= 1$, $\eta = 1$ panel of Fig. \ref{fig:dust_eta}).

For species with $\mathrm{St}= 2$, the density of solid material at the outer part of the corotation region increases when $\mathrm{\eta}= 1$.
Consequently, in the BR case, the region situated behind the planet exerts a greater influence on the solid torque, leading to a reduction {of} the torque ratio to about 117 percent ($\mathrm{St}= 2$, $\mathrm{\eta}= 1$ panel of Fig. \ref{fig:dust_eta}).

For $\mathrm{St}= 3$, the density of solid material at the outer part of the corotation region increases with the accretion efficiency.
However, {for $\mathrm{\eta}= 1$} this change in distribution only {results} in a torque ratio of 97 percent.

For $\mathrm{St}= 4$ and 5 both the density enhancement ($\Lambda$-shape pattern) and depression (pattern behind the planet) {increase} with accretion efficiency.
As a result, the {magnitudes of the solid torque vary} only slightly in {the} BR models.
{Since} the above phenomenon weakens with Stokes number, it is evident that the {effect} of accretion on the torque ratio diminishes with increasing Stokes number for the boulder-sized solid species.

For $\mathrm{St}= 10$, the region around the planet is emptied with increasing accretion efficiency.
However, the torque ratio {decreases by only} about 1 percent from $\eta = 0$ to $\eta = 1$ ($\mathrm{St}= 10$ panels of Fig. \ref{fig:dust_eta}).

\section{Spatial distribution of gas around an Earth-mass planet}
\label{seq:ddistgas}

\begin{figure*}[h!]
\includegraphics[width=17 cm]{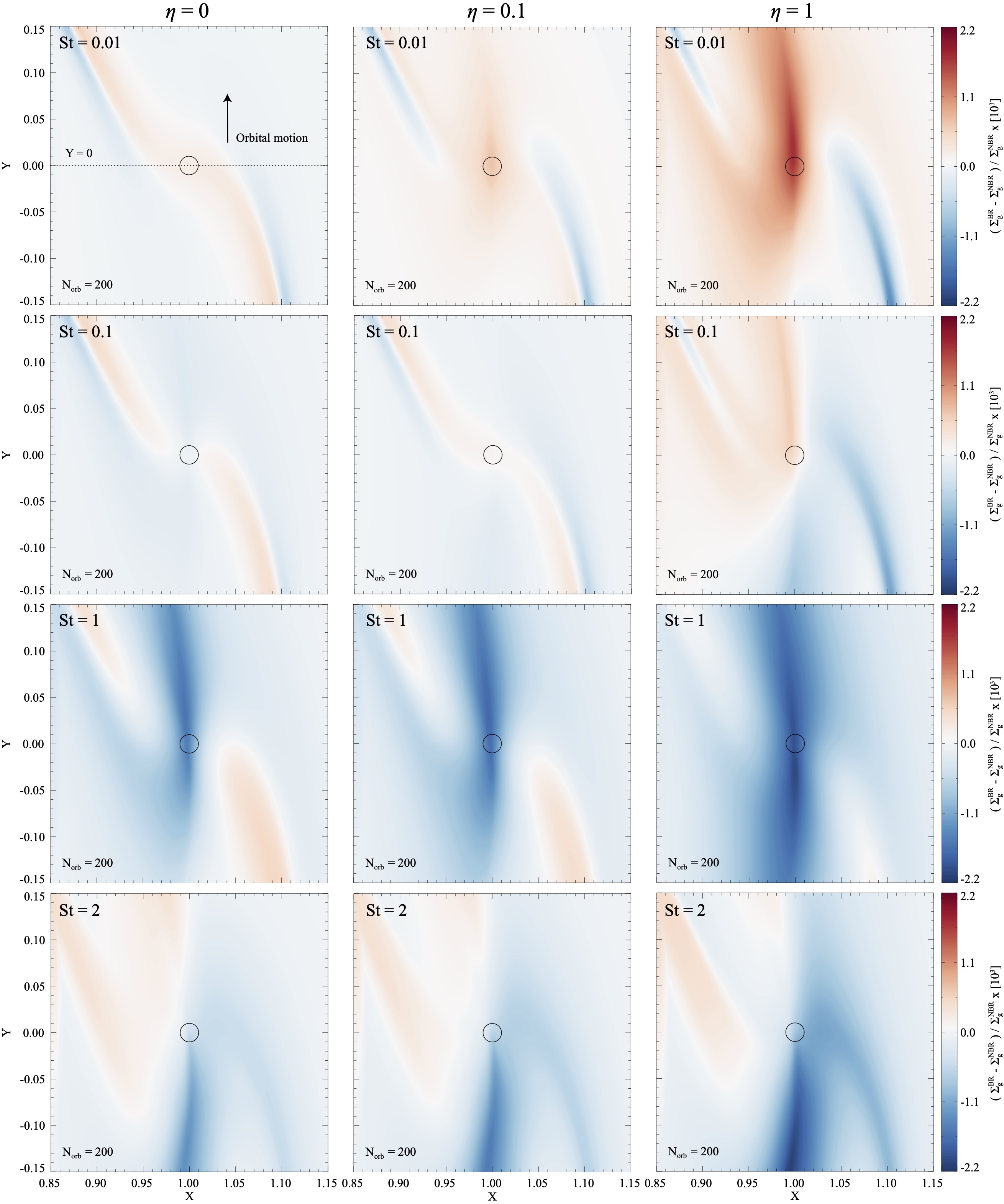}
\centering
\caption{Comparison of gas density distributions in NBR and BR models for $\mathrm{St}= 0.01,~0.1,~1,$ and 2, with three different accretion efficiencies ($\mathrm{\eta}= 0,~0.1,$ and 1), in the vicinity of 1 $M_\oplus$ planet.
The axes are {shown} in code units.
{Colors} ranging from blue to red represent increasing gas densities in {the} BR models. 
Regions where the two models {agree are show} in white. 
The black circle indicates the planetary Hill sphere.
Note that the density ratios are multiplied by 1000 for clarity.
}
\label{fig:com_gas1}
\end{figure*}

\begin{figure*}
\includegraphics[width=17 cm]{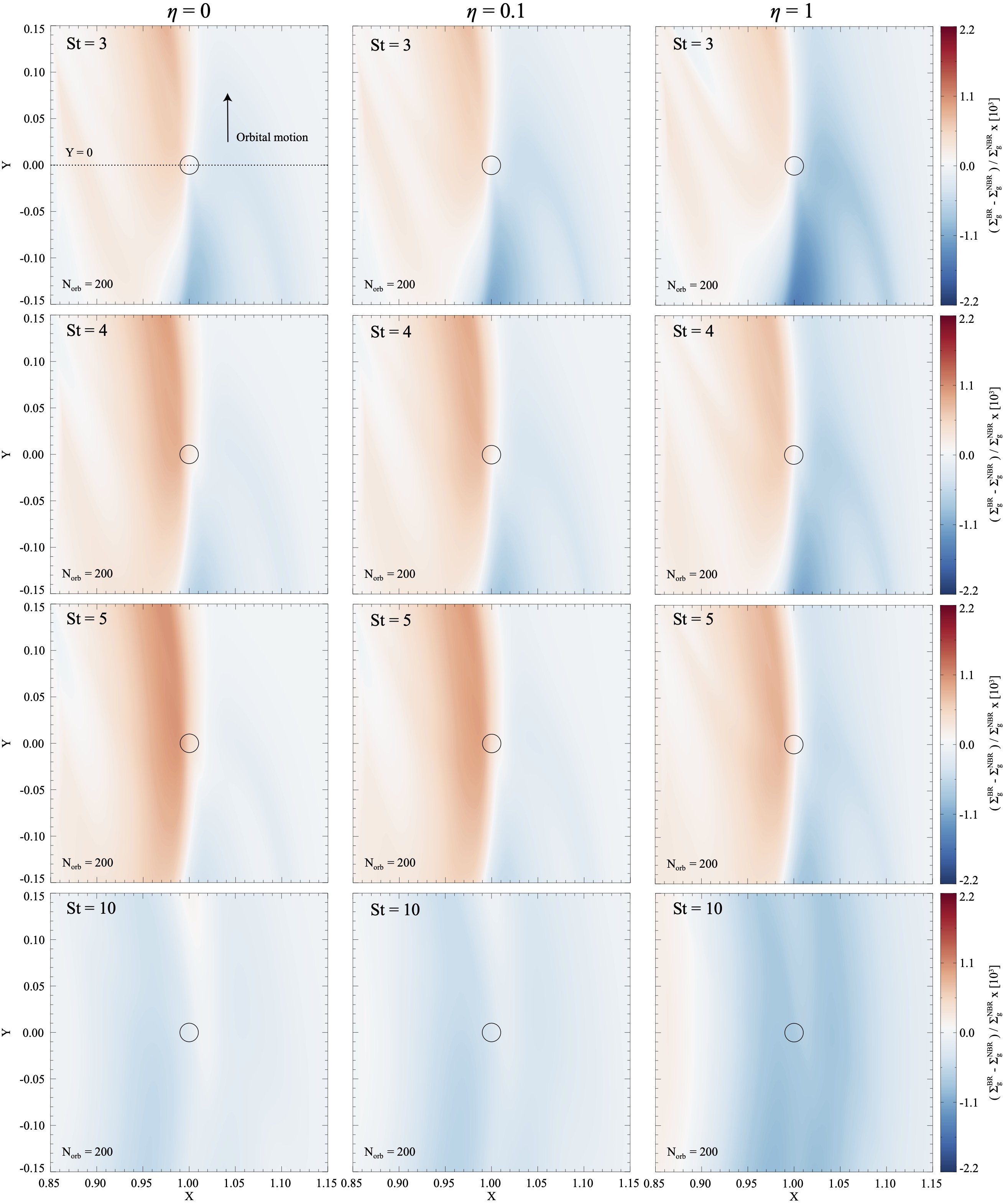}
\centering
\caption{Same as Fig.~\ref{fig:com_gas1} but for $\mathrm{St}= 3,~4,~5,$ and 10.}
\label{fig:com_gas2}
\end{figure*}

Figures~\ref{fig:com_gas1} and \ref{fig:com_gas2} illustrate the comparison of the gas density distribution of NBR and BR models, with three different {efficiencies} of solid accretion, around 1 $M_\oplus$ planet in {the} $p= 0.5$ disk.
The comparison is calculated as $(\Sigma_\mathrm{g}^\mathrm{BR}-\Sigma_\mathrm{g}^\mathrm{NBR})/\Sigma_\mathrm{g}^\mathrm{NBR}$. 
The colors ranging from blue to red represent increasing gas densities in BR models. 
Regions where the two models {agree are shown} in white.
The density distributions are shown in a co-moving frame of the planet.
{Note} that the values {shown} in the figures are multiplied by a factor of $10^{3}$ for the sake of clarity.

First, we {examined} the cases where 
$\mathrm{\eta}= 0$ (left column of Figs.~\ref{fig:com_gas1} and \ref{fig:com_gas2}). 
In the case of the coupled pebble-sized species ($\mathrm{St}=0.01$), an increase in gas density is observed in the BR model within the region where spiral waves have formed as a result of the gravitational interaction between the planet and the disk.
{In addition}, {a decrease in gas density is} observed {both} the {inside} and {outside} the inner and outer spiral {waves}, respectively.
These variations in the distribution only result in a {BR to NBR} torque ratio of about 99 percent, due to the fact that the {variation is symmetric} to {the} $Y=0$ axis   ($\mathrm{St}=0.01$, $\eta=0$ panel of Fig. \ref{fig:gas_eta}).
{A similar pattern is observed for} $\mathrm{St}= 0.1$.
However, the corotation region {ahead of} the planet {shows} a slight {decrease} in the gas in {the} BR model.
As a result, the torque ratio increases, and the planet experiences an increased magnitude {of} negative torque from {the} gas ($\mathrm{St}=0.1$, $\eta=0$ panel of Fig. \ref{fig:gas_eta}).

For species with $\mathrm{St}=1$, the back-reaction also causes  enhanced spiral waves in the gas similar to the coupled species.
However, the gas in the corotation region {shows} a more pronounced depletion, {especially} in the region {in front} of the planet.
As a result, the regions {in front} the planet contribute less to the torque exerted on the planet in the BR case. 
This {explains} the observed {BR to NBR} torque ratio and {the} increase in the magnitude of the negative torque ($\mathrm{St}= 1$, $\eta = 0$ panel of Fig. \ref{fig:gas_eta}).

For $\mathrm{St}= 2$, the back-reaction {effect leads to} a {decrease} of the gas density in the corotation region behind the planet, {compared} to the NBR case.
{In addition}, the density of {the} gas within the planetary orbit and {in front of} the planet, also increases.
Therefore, the contribution to the gas torque from the inner and outer regions is {larger} and smaller, respectively, in the BR case.
This results in a torque ratio of {about} 8 percent and a {significant} reduction in the magnitude of the negative gas torque ($\mathrm{St}= 2$, $\eta = 0$ panel of Fig. \ref{fig:gas_eta}). 

{In the BR case, for species with $\mathrm{St}= 3$ ($\mathrm{St} = 3$, $\eta =0$ panel of Fig. \ref{fig:com_gas2}), the gaseous material inside the planetary orbit increases while the material outside the orbit decreases.}
{Thus}, the effect of the back-reaction of this species {leads} to a markedly asymmetric distribution of {the} gas with respect to the $Y = 0$ axis, thereby generating a {significant} weakening of the magnitude of the negative gas torque in the BR case ($\mathrm{St} = 3$, $\eta =0$ panel of Fig. \ref{fig:gas_eta}).
For $\mathrm{St}= 4-5$, a similar pattern {appears}.
The accumulation of gaseous material inside and outside the planetary orbit is observed to {increase} with the Stokes number.
Consequently,  the {BR to NBR} torque ratio and the magnitude of the negative torque are observed to increase in the BR case when compared to species with $\mathrm{St} = 2-3$.
Nevertheless, the reduction in torque magnitude in BR models remains significant {compared} to the NBR case ($\mathrm{St} = 4-5$, $\eta =0$ panels of Fig. \ref{fig:gas_eta}). 

In the BR model, for species with $\mathrm{St}= 10$, there is a slight decrease and increase in the density of {the} gas density present inside and outside the planetary orbit, respectively.
Therefore, the region {in front of} the planet has a slightly {larger effect} on the gas torque in the BR case, while the regions beyond the planet {have a smaller effect}.
As a result,  the magnitude of the negative torque is reduced in the BR case ($\mathrm{St} = 10$, $\eta = 0$ panel of Fig. \ref{fig:gas_eta}).

The middle and right columns of Figs.~\ref{fig:com_gas1}-\ref{fig:com_gas2} {show} the comparison of the gas density distribution in two {different} accretion regimes, $\mathrm{\eta}= 0.1$ and 1, respectively.
For $\mathrm{St}= 0.01$, as the {accretion efficiency} increases, the amount of gas in the corotation region {in front of} the planet also increases.
Conversely, the density of gas present in the spiral waves is observed to decrease {compared} to the NBR case.
Thus, the region {in front of} the planet {contributes more} to the gas torque exerted on the planet.
As a result, the magnitude of the negative torque decreases with accretion efficiency.
For $\mathrm{St}= 0.1$, the region {in front of} the planet is no longer reduced in gaseous material compared to the NBR case {for} $\eta=0.1$.
As a consequence, the magnitude of the gas torque is observed to decrease in the BR case, resulting in a BR-to-NBR torque ratio of {about} 89 percent ($\mathrm{St} = 0.1$, $\eta =0.1$ panel of Fig. \ref{fig:gas_eta}).
Moreover, in the BR case, high accretion efficiency {leads to} an increase and a decrease in the density of {the} gas inside and outside the orbit of the planet, respectively.
The aforementioned variation in the distribution {leads to} a { noticeable} reduction in the magnitude of the negative gas torque in the BR case ($\mathrm{St} = 0.1$, $\eta =1$ panel of Fig. \ref{fig:gas_eta}).

The back-reaction of species with $\mathrm{St}= 1$ causes the difference in gas distribution of the spiral waves to {decrease} with increasing accretion efficiency.
{In addition}, the {gas density} within the corotation region {continues to decrease and extends} the region behind the planet {for} $\mathrm{\eta}= 0.1$ and 1.
As a result, the magnitude of the gas torque {shows} a slight decrease with increasing accretion efficiency ($\mathrm{St} = 1$, $\eta = 0.1$ and 1 panels of Fig. \ref{fig:gas_eta}).

{For $\mathrm{St}= 2$, the density in the spiral waves behind the planet decreases as the accretion efficiency increases.}
Moreover, the amount of gas inside the planetary orbit increases compared to the NBR case.
{As a result}, the BR-to-NBR torque ratio decreases significantly and even changes sign when $\eta = 0.1$.
For $\eta = 1$, the torque ratio and the magnitude of the positive gas torque {increase} due to solid accretion ($\mathrm{St} = 2$ panels of Fig. \ref{fig:gas_eta}).

For species $\mathrm{St}= 3-5$, solid accretion slightly decreases and increases the amount of gas inside and outside the planetary orbit, respectively.
As a result, the spiral waves preceding and following the planet will have more positive and less negative {contributions} to the gas torque in the BR case, respectively.
This leads to {a} further weakening of the torque ratio between the BR and NBR cases.
{Note} that for $\mathrm{St}= 3$ the gas torque also changes sign when $\eta = 1$ ($\mathrm{St} = 3-5$, $\eta = 0.1,$ and 1 panels of Fig. \ref{fig:gas_eta}).
Thus, the gas distribution becomes increasingly asymmetric with respect to the $Y =0$ axis, which also {leads to} an increase of the BR to NBR torque ratio for these species ($\mathrm{St} = 3-5$, $ \eta = 0.1$ and 1 panels of Fig. \ref{fig:gas_eta}).
{However}, the {increase} in the torque difference is also observed to {decrease} with the Stokes number, analogous to the $\eta = 0$ {scenario}.

For $\mathrm{St}= 10$, {the} accretion {of} solid material {leads to a} slight depletion of gas {near} the planetary orbit in the BR case.
However, these changes result in {only a small} increase in the torque ratio and the magnitude of the negative gas torque does not change significantly ($\mathrm{St} = 10$, $ \eta = 0.1$ and 1 panels of Fig. \ref{fig:gas_eta}).

\end{document}